\documentclass[12pt,preprint]{aastex}

\shorttitle{Savin et al.}
\shortauthors{Dielectronic and Radiative Recombination}

\begin{document}

\title{Dielectronic Recombination (via $N=2 \to N^\prime =2$ Core
Excitations) and Radiative Recombination of \ion{Fe}{20}: Laboratory
Measurements and Theoretical Calculations}

\author{D. W. Savin, E. Behar, and S. M. Kahn}
\affil{Columbia Astrophysics Laboratory and Department of Physics,
Columbia University, \\ New York, NY 10027, USA}
\email{savin@astro.columbia.edu}
\author{G. Gwinner, A. A. Saghiri, M. Schmitt, M. Grieser, R. Repnow,
D. Schwalm, and A. Wolf}
\affil{Max-Planck-Institut f\"{u}r Kernphysik, D-69117 Heidelberg, Germany\\
and Physikalisches Institut der Universit\"at Heidelberg, D-69120
Heidelberg, Germany}
\author{T. Bartsch, A. M\"{u}ller, and S. Schippers}
\affil{Institut f\"{u}r Kernphysik, Strahlenzentrum der
Justus-Liebig-Universit\"{a}t, D-35392 Giessen, \\ Germany}
\author{N. R. Badnell}
\affil{Department of Physics and Applied Physics,
University of Strathclyde, Glasgow, G4 0NG, United Kingdom}
\author{M. H. Chen}
\affil{Lawrence Livermore National Laboratory, Livermore, CA 94550, USA}
\and
\author{T. W. Gorczyca}
\affil{Department of Physics, Western Michigan University,
Kalamazoo, MI 49008, USA}

\begin{abstract} 

We have measured the resonance strengths and energies for dielectronic
recombination (DR) of \ion{Fe}{20} forming \ion{Fe}{19} via $N=2 \to
N^\prime=2$ ($\Delta N=0$) core excitations.  We have also calculated
the DR resonance strengths and energies using AUTOSTRUCTURE, HULLAC,
MCDF, and R-matrix methods, four different state-of-the-art theoretical
techniques.  On average the theoretical resonance strengths agree to
within $\lesssim 10\%$ with experiment.  The AUTOSTRUCTURE, MCDF and
R-matrix results are in better agreement with experiment than are the
HULLAC results.  However, in all cases the $1\sigma$ standard deviation
for the ratios of the theoretical-to-experimental resonance strengths
is $\gtrsim 30\%$ which is significantly larger than the estimated
relative experimental uncertainty of $\lesssim 10\%$.  This suggests
that similar errors exist in the calculated level populations and line
emission spectrum of the recombined ion.  We confirm that theoretical
methods based on inverse-photoionization calculations (e.g., undamped
R-matrix methods) will severely overestimate the strength of the DR
process unless they include the effects of radiation damping.  We also
find that the coupling between the DR and radiative recombination (RR)
channels is small.

Below 2~eV the theoretical resonance energies can be up to
$\approx 30\%$ larger than experiment.  This is larger than the
estimated uncertainty in the experimental energy scale ($\lesssim
0.5\%$ below $\approx 25$~eV and $\lesssim 0.2\%$ for higher energies)
and is attributed to uncertainties in the calculations. These
discrepancies makes DR of \ion{Fe}{20} an excellent case for testing
atomic structure calculations of ions with partially filled shells.
Above 2~eV, agreement between theory and experiment improves
dramatically with the AUTOSTRUCTURE and MCDF results falling within 2\%
of experiment, the R-matrix results within 3\%, and HULLAC within 5\%.
Agreement for all four calculations improves as the resonance energy
increases.

We have used our experimental and theoretical results to produce
Maxwellian-averaged rate coefficients for $\Delta N=0$ DR of
\ion{Fe}{20}.  For $k_BT_e \gtrsim 1$~eV, which includes the predicted
formation temperatures for \ion{Fe}{20} in an optically thin,
low-density photoionized plasma with cosmic abundances, the
experimental and theoretical results agree to better than $\approx
15\%$.  This is within the total estimated experimental uncertainty
limits of $\lesssim 20\%$.  Agreement below $\approx 1$~eV is difficult
to quantify due to current theoretical and experimental limitations.
Agreement with previously published $LS$-coupling rate coefficients is
poor, particularly for $k_BT_e \lesssim 80$~eV.  This is attributed to
errors in the resonance energies of these calculations as well as the
omission of DR via $2p_{1/2} \to 2p_{3/2}$ core excitations.  We have
also used our R-matrix results, topped off using AUTOSTRUCTURE for
RR into $J\ge 25$ levels, to calculate the rate coefficient for RR
of \ion{Fe}{20}.  Our RR results are in good agreement with
previously published calculations.  We find that for temperatures as
low as $k_BT_e \approx 10^{-3}$~eV, DR still dominates over RR for this
system.

\end{abstract}


\section{Introduction}

Low temperature dielectronic recombination (DR) is the dominant
recombination mechanism for most ions in photoionized cosmic plasmas
\citep{Ferl98a}.  Reliably modeling and interpreting spectra from these
plasmas requires accurate low temperature DR rate coefficients.  Of
particular importance are the DR rate coefficients for the iron $L$-shell ions
(\ion{Fe}{17}-\ion{Fe}{24}).  These ions are predicted to play an
important role in determining the thermal structure and line emission of
X-ray photoionized plasmas \citep{Hess97a,Savi99a,Savi00a} which are
predicted to form in the media surrounding accretion powered sources such
as X-ray binaries (XRBs), active galactic nuclei (AGN), and cataclysmic
variables \citep{Kall01a}.

The need for reliable DR data for iron $L$-shell ions has become
particularly urgent with the recent launches of {\it Chandra} and {\it
XMM-Newton}.  These satellites are now providing high-resolution
X-ray spectra from a wide range of X-ray photoionized sources. Examples of
the high quality of the data that these satellites are collecting are
given by the recent {\it Chandra} observations of the XRB Cyg X-3
\citep{Paer00a} and the AGN NGC 3783 \citep{Kasp00a} and the {\it
XMM-Newton} observations of the AGN NGC 1068 \citep{Kink01a} and the
low-mass XRB EXO 0748-67 \citep{Cott01a}. Interpreting the spectra from
these and other photoionized sources will require reliable DR rate
coefficients.

DR is a two-step recombination process that begins when a free electron
approaches an ion, collisionally excites a bound electron of the ion
and is simultaneously captured.  The electron excitation can be labeled
$Nl_j \to N^\prime l^\prime_{j^\prime}$ where $N$ is the principal
quantum number of the core electron, $l$ its orbital angular momentum,
and $j$ its total angular momentum.  This intermediate state, formed by
the simultaneous excitation and capture, may autoionize.  The DR
process is complete when the intermediate state emits a photon which
reduces the total energy of the recombined ion to below its ionization
limit.  Conservation of energy requires that for DR to go forward
$E_k=\Delta E-E_b$. Here $E_k$ is the kinetic energy of the incident
electron, $\Delta E$ the excitation energy of the initially bound
electron, and $E_b$ the binding energy released when the incident
electron is captured onto the excited ion.  Because $\Delta E$ and
$E_b$ are quantized, DR is a resonant process.  DR via $N^\prime=2 \to
N=2$ core excitations (i.e., $\Delta N \equiv N^\prime - N =0$ DR)
generally dominates the DR process for iron $L$-shell ions in
photoionized plasmas \citep{Savi97a,Savi00a}.

To address the need for accurate low temperature DR rate coefficients
for the iron $L$-shell ions, we have initiated a program of
measurements for DR via $2 \to 2$ core excitations using the heavy-ion
Test Storage Ring (TSR) located at the Max-Planck-Institute for Nuclear
Physics in Heidelberg, Germany \citep{Mull97a}.  To date measurements
have been carried out for $\Delta N=0$ DR of \ion{Fe}{18}
\citep{Savi97a,Savi99a}, \ion{Fe}{19} \citep{Savi99a}, \ion{Fe}{20},
\ion{Fe}{21}, and \ion{Fe}{22}.  Here we present our results for
$\Delta N=0$ DR of \ion{Fe}{20} forming \ion{Fe}{19}.  Preliminary
results were presented in \citet{Savi00a}. Results for \ion{Fe}{21} and
\ion{Fe}{22} will be given in future publications.

$\Delta N=0$ DR of nitrogenlike \ion{Fe}{20} can proceed via a number
of intermediate resonance states.  DR occurs when the autoionizing
\ion{Fe}{19} states, produced in the dielectronic capture process,
radiatively stabilize to a bound configuration. Here $\Delta N=0$
captures led to measurable DR resonances for electron-ion collision
energies between 0 and $\approx 105$~eV and involved the following
resonances
\begin{eqnarray}
\label{eq:channels}
{\rm Fe}^{19+}(2s^2 2p^3[^4S^o_{3/2}]) + e^-
\rightarrow \left\{ \begin{array}{ll}
{\rm Fe}^{18+}(2s^2 2p^3[^2D^o_{3/2}]nl) & (n=17,\ldots,\infty)\\
{\rm Fe}^{18+}(2s^2 2p^3[^2D^o_{5/2}]nl) & (n=15,\ldots,\infty)\\
{\rm Fe}^{18+}(2s^2 2p^3[^2P^o_{1/2}]nl) & (n=13,\ldots,\infty)\\
{\rm Fe}^{18+}(2s^2 2p^3[^2P^o_{3/2}]nl) & (n=12,\ldots,\infty)\\
{\rm Fe}^{18+}(2s   2p^4[^4P_{5/2}]nl) & (n=8,\ldots,\infty)\\
{\rm Fe}^{18+}(2s   2p^4[^4P_{3/2}]nl) & (n=7,\ldots,\infty)\\
{\rm Fe}^{18+}(2s   2p^4[^4P_{1/2}]nl) & (n=7,\ldots,\infty)\\
{\rm Fe}^{18+}(2s   2p^4[^2D_{3/2}]nl) & (n=7,\ldots,\infty)\\
{\rm Fe}^{18+}(2s   2p^4[^2D_{5/2}]nl) & (n=7,\ldots,\infty)\\
{\rm Fe}^{18+}(2s   2p^4[^2S_{1/2}]nl) & (n=6,\ldots,\infty)\\
{\rm Fe}^{18+}(2s   2p^4[^2P_{3/2}]nl) & (n=6,\ldots,\infty)\\
{\rm Fe}^{18+}(2s   2p^4[^2P_{1/2}]nl) & (n=6,\ldots,\infty).
%{\rm Fe}^{18+}(2p^5[^2P_{3/2}]nl) & (n=?,\ldots,\infty)\\
%{\rm Fe}^{18+}(2p^5[^2P_{1/2}]nl) & (n=?,\ldots,\infty).
\end{array} \right.
\end{eqnarray}
The lowest lying $\Delta N=1$ resonances are predicted to occur at $E_k
\approx 245$~eV.  The excitation energies $\Delta E$ for all \ion{Fe}{20}
levels in the $n=2$ shell are listed, relative to the ground state, in
Table~\ref{tab:energylevels}.

The experimental technique used here is presented in
\S~\ref{sec:ExperimentalTechnique}.  Our results are given in
\S~\ref{sec:Results}.  Existing and new theoretical calculations are
discussed in \S~\ref{sec:Theory}.  A comparison between theory and our
experimental results is given in \S~\ref{sec:Discussion} and conclusions
in \S~\ref{sec:Conclusions}.

\section{Experimental Technique}
\label{sec:ExperimentalTechnique}

DR measurements are carried out by merging, in one of the straight
sections of TSR, a circulating ion beam with an electron beam. After
demerging, recombined ions are separated from the stored ions using a
dipole magnet and directed onto a detector.  The relative electron-ion
collision energy can be precisely controlled and the recombination
signal measured as a function of this energy. Details of the
experimental setup have been given elsewhere
\citep{Kilg92a,Lamp96a,Savi97a,Savi99a}.  Here we discuss only those
new details of the setup which were specific to our \ion{Fe}{20}
results.

A beam of 280 MeV $^{56}$Fe$^{19+}$ ions was produced and injected into
TSR by the usual techniques.  Stored ion currents of between
$\approx 7-22$~$\mu$A were achieved.  The storage lifetime was $\approx 7$~s.
After injection, the ions were cooled for $\approx 2$~s before data
collection began.  This is long compared to the lifetimes of the various
\ion{Fe}{20} metastable levels (Cheng, Kim, \& Desclaux 1979) and all
ions were assumed to be in their ground state for the
measurements.

The electron beam was adiabatically expanded from a diameter of $\approx
0.95$~cm at the electron gun cathode to $\approx 3.6$~cm before it was merged
with the ions.  In the merged-beams region, the electrons were guided with
a magnetic field of $\approx 40$~mT and traveled co-linear with the stored
ions for a distance of $L\approx 1.5$~m.  The effective energy spread
associated with the relative motion between the ions and the electrons
corresponds to temperatures of $k_BT_\perp \approx 15$~meV perpendicular to
the confining magnetic field and $k_BT_\| \approx 0.13$~meV parallel to the
magnetic field.  The electron density varied between $n_e \approx 1-3 \times
10^7$~cm$^{-3}$.

Data were collected using three different schemes for chopping the
electron beam between the energies for cooling ($E_c$), measurement
($E_m$), and reference ($E_r$).  For center-of-mass collision energies
$E_{cm} \lesssim 0.048$~eV, the chopping pattern (Mode A) began by jumping
to $E_c$ and allowing for a 1.5~ms settling time of the power supplies,
followed by a simultaneous cooling of the ions and collecting of data for
30~ms.  This was followed by a jump to $E_m$, allowing for a 1.5~ms
settling time, and then collecting data for 5~ms.  The pattern was
completed by jumping to $E_r$, allowing for a 1.5~ms settling time, and
then collecting data for 5~ms.  For $E_{cm} \gtrsim 0.048$~eV, two
different chopping patterns were used.  Mode B was similar to Mode A
except that when jumping to $E_m$, a settling time of 20~ms was used, and
data were then collected for 20~ms.  Mode C was similar to Mode B except
an $E_c$-$E_r$-$E_m$ chopping pattern was used.  The chopping pattern was
repeated $\approx 300$ times between injections of new ion current.  With
each step in the chopping pattern, $E_m$ was increased (or decreased) in
the lab frame by $\approx 0.5$~eV.  The electron energy was stepped by this
amount for all three modes.

The reference energy $E_r$ was chosen so that radiative recombination (RR)  
and DR contributed insignificantly to the recombination counts collected
at $E_r$.  This count rate was due to essentially only charge transfer
(CT) of the ion beam off the rest gas in TSR. Taking electron beam
space charge effects into account, the reference energy was $\approx 1600$~eV
greater than the cooling energy of $\approx 2740$~eV.  This corresponds to an
$E_{cm} \approx 183$~eV.

Center-of-mass collision energies were calculated using the velocities of
the electrons and the ions in the overlap region.  The electron velocity
was calculated using the calibrated acceleration voltage and correcting
for the effects of space charge in the electron beam using the beam energy
and diameter and the measured beam current. The ion velocity is determined
by the electron velocity at cooling.

For \ion{Fe}{20}, the DR resonance energies measured using Mode C did
not precisely match those measured using Mode B.  In the lab frame,
resonances measured using Mode C occurred at energies $\approx
1.0-1.5$~eV lower than those using Mode B.  This shift is attributed to
$E_r$ preceding $E_m$ for mode C versus $E_c$ preceding $E_m$ in mode
B.  Capacitances in the electron cooler prevented the acceleration
voltage from reaching the desired value in the time allotted.  For the
data collected here, $E_c$ was essentially always smaller than $E_m$
and $E_r$ was always larger than $E_m$.  Hence in mode B, when the beam
energy was chopped from $E_c$ up to $E_m$, the cooler capacitances
prevented the beam energy from increasing all the way to $E_m$ and the
true electron beam energy was slightly less than expected.  Conversely,
in mode C when the beam energy was chopped from $E_r$ down to $E_m$,
these capacitances prevented the beam energy from decreasing all the
way to $E_m$ and the true beam energy was slightly higher than
expected.  $E_{cm}$ was calculated using the expected electron beam
energy.  Thus the calculated energies in mode B were slightly too high
and in mode C slightly too low.  To merge the Mode B and Mode C data
sets we shifted the Mode C data up in energy, in the lab frame, by
$\approx 1.0$~eV at moderate energies and $\approx 1.5$~eV at higher
energies.  Technical reasons for the occurrence of these voltage errors
have been identified and corrected. 

The systematic inaccuracies in the absolute $E_{cm}$ scale derived from
the voltage calibrations were $\lesssim 2\%$.  To increase the
accuracy of the $E_{cm}$ scale, a final normalization of the $E_{cm}$
scale was performed using calculated energies for the DR resonances,
\begin{equation}
\label{eq:e_nl}
E_{nl} = \Delta E - \Biggl({z\over n-\mu_l}\Biggr)^2 {\cal R}.
\end{equation}
Here $E_{nl}$ is the resonance energy for DR into a given $nl$ level, $z$
the charge of the ion before DR, $\mu_l$ the quantum defect for
the recombined ion, and ${\cal R}$ the Rydberg energy.  Values for $\Delta
E$ were taken from spectroscopic measurements \citep{Suga85a} as listed in
Table~\ref{tab:energylevels}.  The quantum defects account for energy
shifts of those $l$ levels which have a significant overlap with the ion
core and cannot be described using the uncorrected Rydberg formula.  
As $l$ increases, the overlap with the ion core decreases and $\mu_l$ goes
to zero.

For the normalization of the $E_{cm}$ scale we used DR resonances with
$n\ge 7$ which were essentially unblended with other resonances.  We
considered only the high-$l$ contributions occurring at the highest
energy of a given $n$ manifold, for which $\mu_l$ is essentially zero.
The resulting calculated resonance energies were $\approx 1.046$ times
the experimental energy scale for $E_{cm} \approx 0.17$~eV.  This
factor decreased nonlinearly with increasing energy to $\approx 1.016$
at $\approx 10$~eV and then slowly decreased to $\approx 1.003$ with
increasing energy.  We multiplied the experimental energy scale by this
energy-dependent normalization factor to produce the final energy scale
for the results presented here.  After corrections, we estimate that
above $\approx 25$~eV, the uncertainty in the corrected energy scale is
$\lesssim 0.2\%$.  Below $\approx 25$~eV, it is estimated to be
$\lesssim 0.5\%$.

The electron and ion beams were merged and then, after passing
through the interaction region, they were separated using toroidal
magnets.  The motional electric fields in the downstream toroidal
magnet field-ionized electrons which had dielectronically recombined
into Rydberg levels $n\gtrsim n_{cut1} = 146$.  Further downstream, two
correction dipole magnets field-ionized electrons in levels $n\gtrsim
n_{cut2} = 120$.  Finally, the recombined ions passed through a dipole
which separated them from the primary ion beam and directed them onto a
detector.  Electrons in $n \gtrsim n_{cut3} = 64$ were field ionized by
this magnet.  The flight time of the ions from the center of the
interaction region to the final dipole magnet was $\approx 166$~ns.
During this time some of the captured electrons radiatively decayed
below the various values of $n_{cut}$.  DR occurs primarily into $l
\lesssim 8$ levels.  Using the hydrogenic formula for radiative
lifetimes of \citet{Marx91a}, we estimate that for DR into $n \lesssim
n_{max} = 120$, the captured electrons radiatively decayed below the
various values of $n_{cut}$ before reaching the final dipole and were
therefore detected by our experimental arrangement.

The measured recombination signal rate was calculated by taking the
rate at the measurement energy $R(E_{cm})$ and subtracting from it the
corresponding rate at the reference energy $R(E_{ref})$.  This
eliminates the effects of slow pressure variations during the scanning
of the measurement energy but not the effects of any fast pressure
variations associated with the chopping of the electron beam energy,
leaving a small residual CT background.  Following \citet{Schi01a},
the measured rate coefficient $\alpha(E_{cm})$ is given by 
\begin{equation}
\label{eq:rate}
\alpha_L(E_{cm})= {[R(E_{cm})-R(E_{ref})]\gamma^2 \over
n_e N_i(L/C)\eta} + \alpha(E_{ref}) { n_e(E_{ref}) \over n_e(E_{cm})}.
\end{equation}
Here $N_i$ is the number of ions stored in the ring, $C=55.4$~m the
circumference of the ring, $\eta$ the detection efficiency of the
recombined ions (which is essentially 1), $\gamma^2=[1-(v/c)^2]^{-1}
\approx 1.01$, and $c$ the speed of light.  The measured rate
coefficient represents the DR and RR cross sections multiplied by the
relative electron-ion velocity and then convolved with the experimental
energy spread. The data sit on top of the residual CT background.  The
experimental energy spread is best described by an anisotropic
Maxwellian distribution in the comoving frame of the electron beam.
The second term in Equation~\ref{eq:rate} is a small correction to
re-add the RR signal at the reference which is subtracted out in the
expression $[R(E_{cm})-R(E_{ref})]$.  Here we used the theoretical RR
rate coefficient at $E_{cm} = 183$~eV where contributions due to DR are
insignificant.  The RR rate coefficient
at this energy, calculated using a modified
semi-classical formula for the RR cross section \citep{Schi98a}, is
$\approx 4.3\times 10^{-12}$ cm$^3$ s$^{-1}$. Using $\alpha_L(E_{cm})$,
the effects of the merging and demerging of the electron and ion beams
are accounted for, following the procedure described in
\citet{Lamp96a}, to produce a final measured recombination rate
coefficient $\alpha(E_{cm})$ from which the DR results are extracted.

The DR resonances produce peaks in $\alpha(E_{cm})$.  Resonance
strengths are extracted after subtracting out the smooth background due
to RR and CT.  Although RR dominates the smooth background at low
energies, we have been unable to extract reliable RR rate coefficients
due to the remaining CT contributions to the measured signal rate.

Experimental uncertainties have been discussed in detail elsewhere
\citep{Kilg92a,Lamp96a}. The total systematic uncertainty in our
absolute DR measurements is estimated to be $\lesssim 20\%$.  The major
sources of uncertainties include the electron beam density
determination, the ion current measurement, corrections for the merging
and demerging of the two beams, the efficiency of the recombined ion
detector, resonance strength fitting uncertainties, and uncertainties
in the shape of the interpolated smooth background (particularly in
regions where the DR resonances were so numerous that the background
was not directly observable).  Another source of uncertainty is that we
assume each DR feature can be fit using a single resonance peak when in
fact each feature is often composed of many unresolved resonance
peaks.  Relative uncertainties for comparing our DR results at
different energies are estimated to be $\lesssim 10\%$.  Uncertainties
are quoted at a confidence level believed to be equivalent to a 90\%
counting statistics confidence level.

\section{Experimental Results}
\label{sec:Results}

Our measured spectrum of \ion{Fe}{20} to \ion{Fe}{19} $\Delta N=0$ DR
resonances is shown in Figure~\ref{fig:FeXXresonances}(a).  The data
represent the sum of the RR and DR cross sections times the relative
electron-ion velocity convolved with the energy spread of the
experiment, i.e., a rate coefficient.  The data are presented as a
function of $E_{cm}$.  For energies below
7.5~eV, we use the predicted asymmetric line shape for the DR
resonances \citep{Kilg92a} and fit the data to extract DR resonance
strengths and energies. Above 7.5~eV, the asymmetry is insignificant
and we fit the data using Gaussian line shapes.  Extracted resonance
strengths $S_d$ and energies $E_d$ for a given DR resonance or blend of
resonances $d$ are listed in Table~\ref{tab:FeXXextracteddata}.  The
energies have been corrected as described in
\S~\ref{sec:ExperimentalTechnique}.

The lowest-energy resolved resonance is the $2s^2 2p^3 (^2D^o_{3/2})
17l$ blend at $E_{cm} \approx 0.081$~eV.  Our fit to this blend begins
to deviate significantly from the measured data for $E_{cm} \lesssim
0.05$~eV (see Figure~\ref{fig:nearzero}).  We attribute this deviation
to unresolved broad and narrow DR resonances lying below 0.05~eV.

Due to the energy spread of the electron beam, resonances below $E_{cm}
\approx k_BT_e \approx 0.015$~eV cannot be resolved from the near 0~eV
RR signal.  However, we can infer the presence of such resonances.  The
measured recombination rate coefficient at $E_{cm} \lesssim 10^{-4}$~eV
is a factor of $\approx 90$ times larger than the RR rate coefficient
predicted using semiclassical RR theory with quantum mechanical
corrections \citep{Schi98a}.  This enhancement factor is much larger
than that found for \ion{Fe}{18} for which the near 0~eV recombination
rate coefficient was a factor of $\approx 2.9$ times larger than the
theoretical RR rate coefficient.  \ion{Fe}{18} is predicted to have no
DR resonances near 0~eV.  A similar enhancement (factor of $\approx
2.2$) was found for RR of bare \ion{Cl}{18} \citep{Hoff01a}.  For
\ion{Fe}{19}, the enhancement was a factor of $\approx 10$.
\ion{Fe}{19} and \ion{Fe}{20} are both predicted to have near 0~eV DR
resonances and the inferred enhancement factors of greater than 2.9 are
attributed to these unresolved near 0~eV resonances.

We note that a number of issues pertaining to recombination
measurements in electron coolers at $E_{cm} \lesssim k_BT_e$ remain to
be resolved \citep{Hoff98a,Schi98a,Gwin00a,Hoff01a}, but it is highly
unlikely that their resolution will lead to a near 0~eV recombination
rate coefficient that increases by a factor of $\approx 30$ for a
change in ionic charge from 17 to 19.  Thus we infer that there are
unresolved DR resonances lying at energies below 0.015~eV.

Our calculations suggest that these unresolved resonances are due to a
combination of the $2s^22p^3(^2D^o_{5/2})15l$ and $2s 2p^4 (^4P_{3/2})
7d$ configurations.  Calculations indicate these $15l$ resonances have
natural line widths significantly smaller than the energy spread of the
experiment.  Here we treat them as delta functions for fitting
purposes.  To determine the energies of these $15l$ resonances, we use
the calculated quantum defect for an $nf$ electron in \ion{Fe}{19} from
\citet{Theo86a}.  The $f$ level is the highest angular momentum they
considered.  We extrapolate this quantum defect to higher angular
momentum using the predicted $l^{-1}$ behavior \citep{Babb92a}.  The
resulting resonance energies are listed in
Table~\ref{tab:FeXXextracteddata}.  We estimate that for this complex,
the $15i$ level is the lowest lying DR resonance.  The highest
resonance energy (for the $15t$ level) is estimated to be at $\approx
0.005$~eV.

The energy of the near 0~eV $2s 2p^4 (^4P_{3/2}) 7d$ resonance is
difficult to predict reliably because of the large interaction of the
captured electron with the core.  Calculations indicate the resonance
has a width of $\approx 10$~meV which is comparable to the energy
spread of the experiment.  To fit for this feature we must take the
natural line profile of the DR resonance and its $E^{-1}_{cm}$
dependence into account.  \citet{Mitn99a} have addressed theoretically
the issue of near 0~eV DR resonances.  Starting from Equation~12 of
their paper, we can write the near 0~eV DR line profile as
\begin{equation}
\sigma_{DR}^d(E_{cm})= 
{S_d E_d \over E_{cm}}
\Biggr[{\Gamma_d/2\pi \over (E_{cm}-E_d)^2 + (\Gamma_d/2)^2} \Biggr]
\label{eq:drprofile}
\end{equation}
where $\Gamma_d$ is the natural line width of the resonance.

Recent measurements of recombination of bare \ion{Cl}{18} found an
enhanced recombination rate coefficient for $E_{cm} \lesssim 0.008$~eV
\citep{Hoff01a}.  We expect a similar situation for \ion{Fe}{20}.
Because the unresolved $15l$ DR resonances all occur for $E_{cm} \lesssim
0.005$~eV, we attribute the DR signal between 0.008 and 0.05~eV to
the unresolved $7d$ resonance.  We have fit this portion of the
recombination spectrum essentially by eye, varying the resonance width,
strength, and energy.  Our best fit was for an inferred resonance width
of 10 meV.  The inferred resonance energy and strength of this $7d$
resonances are listed in Table~\ref{tab:FeXXextracteddata}.

Based on our \ion{Fe}{18} results \citep{Savi97a,Savi99a}, we expect to
see an enhancement of $\approx 2.9$ as $E_{cm}$ approaches 0~eV.
Taking only the near 0~eV $7d$ resonance into account yields an
enhancement factor of $\approx 6.7$.  We infer the resonance strength
of the near 0~eV $15l$ resonances by varying their amplitudes to
produce a model recombination spectrum which yields an enhancement
factor of $\approx 2.9$.

We have linked the resonance strengths of the near 0~eV $15l$ levels
taking into account the behavior of the DR cross section.  Following
the logic in \S~II of \citet{Mull87a}, when the radiative stabilization
rate $A_r$ is much greater than the autoionization rate $A_a$ of the
intermediate doubly-excited state in the DR process, then the DR
resonance strength is proportional to $A_a$.  For the
$2s^22p^3(^2D^o_{5/2})15l$, the excited core electron cannot decay via
an electric dipole transition.  Stabilization of the intermediate
autoionizing state is due to a radiative decay by the Rydberg electron.
Using the hydrogenic formula of \citet{Marx91a} for the radiative
lifetime of the $15l$ electron and our calculated MCDF autoionization
rates, we find that the radiative rates are always significantly larger
than the autoionization rates.  We have therefore linked the relative
resonance strengths for the near 0~eV $15l$ resonances using the MCDF
calculated $A_a$ values.  Thus the amplitudes of these resonances are
controlled by a single normalization factor.  We have varied this
factor until our model recombination spectrum yields an enhancement
factor of $\approx 2.9$ for $E_{cm} < 10^{-4}$~eV.  The inferred resonance
strengths for these $15l$ resonances are listed in
Table~\ref{tab:FeXXextracteddata}.

The measured and model recombination spectrum below $E_{cm}=0.1$~eV is
shown in Figure~\ref{fig:nearzero}.  For the model spectrum we use our
inferred and extracted resonance strengths and energies.  We have
looked at the difference between the measured and model spectrum
between 0.008 and 0.05~eV.  The resulting residuals are comparable to
the difference between the measured spectrum and the fitted spectrum
for those peaks below 1~eV which we were able to fit using a $\chi^2$
procedure.  We note here that the 10~meV width of this resonances is
significantly larger than our fitted resonance energy of 3~meV.  Thus
we infer that the DR cross section is non-zero in value for
$E_{cm}=0$~eV and that the resulting Maxwellian DR rate coefficient
will increase as the plasma temperature decreases.

We have used the extracted DR resonance strengths and energies listed
in Table~\ref{tab:FeXXextracteddata} to produce a rate coefficient for
$\Delta N=0$ DR of \ion{Fe}{20} forming \ion{Fe}{19} in a plasma with a
Maxwellian electron energy distribution at a temperature $T_e$.  We
treated all resonances listed, except for the near 0~eV $7d$ resonance,
as delta functions.  Using these resonances and the measured unresolved
resonances near the series limit, we have produced a rate coefficient
following the procedure described in \citet{Savi99b}.  To this we have
added the rate coefficient due to the $7d$ resonance.  This rate
coefficient is calculated using Equation~\ref{eq:drprofile} multiplied
by the relative electron-ion velocity and integrating this over a
Maxwellian distribution.  The resulting $\Delta N=0$ rate coefficient
is shown in Figure~\ref{fig:FeXXrates}(a).  The inferred contribution
due to the near 0~eV $15l$ and $7d$ resonances is $\approx 81\%$ at
$k_BT_e = 0.1$~eV, $\approx 18\%$ at 1~eV, $\approx 4\%$ at 10~eV, and
$\approx 1\%$ at 100~eV.  We estimate the uncertainty in our
experimentally-derived rate coefficient to be $\lesssim 20\%$ for
$k_BT_e \gtrsim 1$~eV.  At lower temperatures, the uncertainty of the
strengths for the near 0~eV resonances causes a larger uncertainty
which is is difficult to quantify.

We have fitted our experimentally-derived $\Delta N=0$ DR rate 
coefficient using
\begin{equation}
\alpha_{DR}(T_e)=
T_e^{-3/2}\sum_i c_i e^{-E_i/k_BT_e}
\label{eq:drratefit}
\end{equation}
where $T_e$ is given in units of K.  Table~\ref{tab:fitparameters}
lists the best-fit values for the fit parameters.  The fit is good to
better than 1.5\% for $0.001 \le k_BT_e \le 10000$~eV.  Although we
infer above that the DR rate coefficient is non-zero at $k_BT_e=0$~eV,
our fitted DR rate coefficient eventually goes to 0 for $k_BT_e <
0.001$~eV.  However, we expect this to have no significant effect on
plasma modeling as it is extremely unlikely that \ion{Fe}{20} will ever
form at temperatures below 0.001~eV \citep{Kall01a}.

\section{Theory}
\label{sec:Theory}

Existing theoretical rate coefficients for DR of \ion{Fe}{20} have
been calculated in $LS$-coupling.
\citet{Shul82a} present the fitted results of \citet{Jaco77a}.
\citet{Arna92a} present the unpublished results of Roszman.  Details
of the theoretical techniques used for the calculations can be found in
\citet{Jaco77a} and \citet{Rosz87a} and references therein.

There have been major theoretical advances in the study of DR since the
works of Jacobs et al.\ and Roszman.  We have carried out new
calculations using AUTOSTRUCTURE, HULLAC, MCDF, and R-matrix methods,
four different state-of-the-art theoretical techniques.  Below we
briefly describe these techniques and the results.

\subsection{AUTOSTRUCTURE}
\label{secauto}

DR cross section calculations were carried out in the
independent-processes, isolated-resonance approximation using the code
AUTOSTRUCTURE \citep{bad86}.  This technique treats both the
electron-electron (repulsive Coulomb) operator $V=\sum_{\alpha\beta}
{1\over \vert \vec{r}_\alpha-\vec{r}_\beta\vert}$ and the
electron-photon (electric dipole) operator
$\vec{D}=\sqrt{2\omega^3\over 3\pi c^3}\sum_\alpha \vec{r}_\alpha$ to
first order.  The subscripts $\alpha$ and $\beta$ are electron labels
and $\omega$ is the emitted photon energy.

All continuum wavefunctions $2l^5\epsilon l^\prime$, and all resonance
or bound wavefunctions $2l^5 n l^\prime$, were constructed within the
distorted-wave approximation.  The resulting wavefunctions were used to
calculate all autoionization rates $\Gamma^a_{di}=2\pi\vert\langle
2l_d^5n_dl^\prime_d\vert V \vert
2l_i^5\epsilon_il_i^\prime\rangle\vert^2$ and radiative rates
$\Gamma^r_{df}=2\pi\vert\langle 2l_d^5n_dl^\prime_d\vert \vec{D} \vert
2l_f^5n_fl_f^\prime\rangle\vert^2$.  Here the subscript $i$ denotes the
continuum states ($i=1$ is the initial free electron plus the initial
ionic system), $d$ denotes the resonance states, and $f$ denotes the
final recombined states.  Next, these rates were all used in the
analytic expression for the (unconvoluted) DR cross section
\begin{eqnarray}
\sigma_{DR}(E)=\sum_d \sigma^d_{DR}(E)=\sum_d {2\pi^2\over k^2}
{(2J^t_d+1)\over 2(2J_{core}+1)}\Gamma^a_{d1}
\left[\sum_{f^\prime}\Gamma^r_{df^\prime}/2\pi
\over (E-E_d)^2+
\left({\sum_i\Gamma^a_{di}+\sum_f\Gamma^r_{df}\over 2}\right)^2\right]\,
\label{eqdr}
\end{eqnarray}
which is a function of electron kinetic energy $E={1\over 2}k^2$
relative to the initial state (e.g.,  $i=1$).  $J^t_d$ is the
total angular momentum of the resonance state, $J_{core}=3/2$ the
angular momentum of the $1s^22s^22p^3(^4S_{3/2})$ initial core ionic
state, and $E_d$ the energy of the resonance state.  The continuum
wavefunctions are energy normalized such that $\langle\epsilon
l\vert\epsilon^\prime l^\prime\rangle
=\delta(\epsilon-\epsilon^\prime)\delta_{ll^\prime}$.  The sum
over $f^\prime$ in the numerator only includes radiative transitions to
bound states.  Radiative decay to states that subsequently autoionize
make rather small contributions to the DR process and are only included in
the sum over $f$ in the denominator.

For the initial atomic structure, the $1s$, $2s$, and $2p$ orbitals
making up all possible $2l^5(^{2S+1}L_J)$ ionic states, as well as the
$2l^6$ recombined states, were determined from a Hartree-Fock
\citep{ff91} calculation for the $1s^22s^22p^3(^4S)$ ground state of
\ion{Fe}{20}.  The 7 and 8 electron atomic structures were obtained by
diagonalizing the appropriate Breit-Pauli Hamiltonian.  Calculated
ionic \ion{Fe}{20} energies are listed in
Table~\ref{tab:energylevels}.  Prior to the final DR cross section
calculations, these ionic thresholds were shifted to the known
spectroscopic values \citep{Suga85a} by $\lesssim 2.5$~eV.  The
$\epsilon_il_i^\prime$ and $n_fl_f^\prime$ orbitals were subsequently
determined from single-configuration continuum and bound distorted wave
calculations, respectively.  We included explicitly all orbital angular
momentum and principal quantum numbers in the range $0\le
l^\prime\le17$ and $6\le n\le 120$.  Configuration mixing was minimal
in these calculations.  Only the $2l^6$ bound states were coupled to
each other.  All other $2l^5nl^\prime$ resonances, for all $n>6$ and
$l^\prime$, were treated as non-interacting resonances.

The DR cross section is the sum of Lorentzian profiles.  This
analytic cross section can also be energy integrated to give resonance
strengths or convoluted with the experimental energy distribution for
comparison with the measured results.  DR rate coefficients
can be obtained by convolving the DR cross section with a
Maxwellian electron distribution.

\subsection{HULLAC}

DR resonance strengths are calculated in the independent processes,
isolated resonance, and low-density approximations. The DR cross
section can then be written as the product of the cross section for
dielectronic capture and the branching ratio for subsequent radiative
stabilization. In the low-density limit, the branching ratio includes
only radiative and autoionization decays. Basic atomic quantities
are obtained using the multi-configuration HULLAC (Hebrew University
Lawrence Livermore Atomic Code) computer package 
\citep{BarS01a}. The calculations employ a relativistic parametric
potential method for the atomic energy levels \citep{Klap71a,Klap77a}
while using first order perturbation theory for the radiative decay
rates. The autoionization rates are calculated in the distorted wave
approximation, implementing a highly efficient
factorization-interpolation method \citep{BarS88a,Oreg91a}.  Full
configuration mixing is included within and between the configuration
complexes $1s^2 2l^5 n^\prime l^\prime (n^\prime \le 6)$.  For the
$1s^2 2l^5 n^\prime l^\prime (n^\prime > 6)$ complexes, only mixings
within a given $n^\prime$-complex are included .  Mixing between
complexes with different $n^\prime$ values for $n^\prime > 6$ has only
a minor effect and is neglected.

All of the dielectronic capture channels from the \ion{Fe}{20}
ground level $1s^2
2s^2 2p^3\ ^4S^o_{3/2}$ to the \ion{Fe}{19}
doubly excited levels $1s^2 2l^5 n'l'$ are
included. These include the fine-structure core excitations (i.e.,
$2p_{1/2}-2p_{3/2}$ core transitions).  Explicit calculations are
performed for $6\le n' \le 25$, and $l'\le 9$.  DR contributions from
$1s^2 2l^5 n'l' (n' > 25)$ configurations are estimated by applying the
$n'^{-3}$ scaling law to the individual autoionization and radiative
transition rates when the $n^\prime$ electron is involved.  Calculated
\ion{Fe}{20} energy levels are listed in Table~\ref{tab:energylevels}.
These correspond to the various series limit energies for $\Delta N=0$
DR.  Prior to the final DR cross section calculations, the theoretical
resonance energies have been adjusted by $\lesssim 2.1$ eV so that the
series limits match the spectroscopically measured energies
\citep{Suga85a}.  All possible autoionization processes to $1s^2 2l^5$
levels following the initial dielectronic capture are accounted for,
including those to excited states.  All of the radiative decays to
non-autoionizing levels are included in the branching ratio.  Radiative
cascades to autoionizing levels, on the average, can be shown to have
little effect on the calculated branching ratios
\citep{Beha95a,Beha96a}.  Throughout this work only the electric dipole
radiative transitions are computed.  The calculated DR cross sections
are folded with a Maxwellian distribution of the plasma electrons to
obtain the DR rate coefficients.

\subsection{Multiconfiguration Dirac-Fock (MCDF)}

DR calculations are carried out in the independent process, isolated
resonance approximation \citep{Seat76a}.  In these approximations, the
interference between DR and RR is neglected and the effects of
interacting resonances are ignored. The DR cross section can then be
written as a product of the resonance capture cross section and the
stabilizing radiative branching ratio.  The required energy levels and 
Auger and radiative transition rates for the autoionizing states are
obtained using the Multiconfiguration Dirac-Fock (MCDF) method
\citep{Gran80a,Chen85a}.  These calculations are carried out in the
average-level scheme and in intermediate coupling with configuration
interaction within the same principal quantum $n$ complex. All possible
Coster-Kronig channels and radiative decays to bound states are
included. A one-step cascade correction is taken into account when the
radiative decay of the core electron leads to an autoionizing state.

We include excitation from the ground state $1s^2 2s^2 2p^3\ ^4S_{3/2}$
to the $1s^2 2s^2 2p^3\  ^2P$, $^2D$ and $1s^2 2s 2p^4\ ^4P$, $^2D$,
$^2S$, and $^2P$ states. For fine-structure core excitations (i.e.,
$2p_{1/2}-2p_{3/2}$ core transitions), explicit calculations are
performed for $12\le n \le 35$, and $l\le12$ autoionizing states. For
$2s-2p$ core excitations, explicit calculations are carried out for
$6\le n \le 35$, and $l\le 12$ states.   Contributions from $l>12$
have been estimated by extrapolating from the $l=10-12$ results.
The contributions 
contribute $<1\%$ to the total DR rate coefficient and are neglected in the final
calculations.  Calculated
\ion{Fe}{20} energy levels are listed in Table~\ref{tab:energylevels}.
These correspond to the various series limit energies for $\Delta N=0$
DR.  Prior to the final DR cross section calculations, the theoretical
resonance energies have been adjusted by $\lesssim 1.5$ eV so that the
series limits match the spectroscopically determined excitation
energies \citep{Suga85a}.  The DR cross sections for $36 \le n \le 120$
states are estimated by using the $n^{-3}$ scaling law for the
transition rates.  DR cross sections with $6 \le n \le 120$ have
been folded with the Maxwellian distribution of the plasma electrons to
obtain the DR rate coefficients.

\subsection{R-Matrix}
\label{secrmat}

We have also carried out calculations using the Belfast R-matrix codes
for the inner region \citep{burke93, berr95} and a modified version of
the STGF code for the outer region \citep{Berr87a}.  These include
spin-orbit and other Breit-Pauli corrections \citep{scott82}, and have
been extensively modified to include radiation damping
\citep{robicheaux95, gorczyca95, gorczyca96}, which is crucial for the
present case of \ion{Fe}{20}.  One appealing aspect of the R-Matrix
technique is that the continua and resonances are coupled together as a
structured continuum, unlike the perturbative methods that compute
resonance and continuum distorted wave orbitals separately.  This is
achieved somewhat differently depending on the region of configuration
space.  Inside the so-called R-matrix ``box'' the total 8 electron
wavefunction of \ion{Fe}{19} is expanded in a large basis, making no
distinction between resonance or continuum states.  The surface
amplitudes at $r_a$, compactly represented by the R-matrix, are
determined from variational considerations.  The radius of the ``box''
used here, $r_a=2.2$ a.u., was chosen in order to include all $2p^53l$
bound states.  Outside the R-matrix box, the continua and resonances
are initially treated as separate Coulomb functions, but are then
coupled by the long-range non-Coulombic potential, giving off-diagonal
elements to the open-closed scattering matrix of multi-channel quantum
defect theory (MQDT).  Thus, the outer region wavefunction is also made
up of structured continua, once physical boundary conditions are
applied.  Note that we find the long-range coupling to significantly
affect the calculated DR cross section \citep{gorczyca96}.

In order to describe how the subsequent radiation from these structured
continua are included in the present treatment, it helps to first show 
all included direct (RR) and resonant (DR) pathways leading to recombination
for the case of \ion{Fe}{20}:
\begin{eqnarray}
e^-+2s^22p^3(^4S) 
 &  \rightarrow & \rightarrow 2s^22p^3(^4S_{3/2})n^\prime l^\prime\
(2\le n^\prime\le3) \label{path1}\\
 & \rightarrow 2l^5 nl & \rightarrow 2l^5n^\prime l^\prime\
(2\le n^\prime \le3) \label{path2}\\
 &  \rightarrow & \rightarrow 2s^22p^3(^4S_{3/2})n^\prime l^\prime\
(4\le n^\prime\le 120) \label{path3}\\
 & \rightarrow 2s2p^4nl& \rightarrow 2s^22p^3nl\
(6\le n\le 120) \label{path4}\\
 & \rightarrow 2s2p^4nl & \rightarrow 2s2p^{4}n^\prime l\pm 1\
(4\le n^\prime\lesssim 5)\ \label{path5}\\
 & \rightarrow 2s^22p^{3*}nl & \rightarrow 2s^22p^{3*}n^\prime l\pm 1\
(4\le n^\prime\lesssim 16). \label{path6}
\end{eqnarray}
In the above pathways,
the stabilizing photon emitted 
has
been omitted.  In Equation~\ref{path4}, the
$2s2p^4nl \to 2s^2 2p^3nl$ radiative transition may leave the
core in either its ground state or an excited state.  
In Equations~\ref{path5} and \ref{path6}, the $\lesssim$ symbols  
indicate that the
exact maximum value of $n^\prime$ depends on the specific configuration
of the core electrons.  This value of $n^\prime$ can be determined from
Equation~\ref{eq:channels} for the different core configurations.  The
notation $2p^{3*}$ indicates that the $2p^3$ electrons are in an
excited configuration.

The direct/resonant processes in Equations \ref{path1} and \ref{path2},
end up in recombined states that reside completely in the R-matrix
box.  Recombination into these states is treated by using a non-local,
energy-dependent, imaginary optical potential in the inner-region
Hamiltonian, leading to a complex R-matrix, and therefore a non-unitary
S-matrix.  Thus, interference between DR and RR is naturally included
here.  For the direct recombination shown in Equation \ref{path3}, we
add a term $-i\Gamma_{RR}/2$ to the diagonal open-open elements of the
scattering matrix, where $\Gamma_{RR}$ is computed in the hydrogenic
approximation as
\begin{eqnarray}
\Gamma_{RR}=2\pi\sum_{n^\prime=4}^\infty\sum_{l,l^\prime} \vert\langle \epsilon l\vert D\vert n^\prime
l^\prime\rangle\vert^2  
\end{eqnarray}
where $\epsilon l$ denotes a continuum orbital.

The RR processes in Equations \ref{path1} and \ref{path3} are also used
to compute a pure RR cross section, but it is important to omit all
excited states $2l^5$ and scatter from the $2s^22p^3(^4S_{3/2})$ target
alone, thereby eliminating all DR resonances.  Here we used partial
waves $J^\pi$ from $J_{max}=10$ to $J_{max}=25$, for both even and odd
parities $\pi$.  In order to get reasonable agreement with the RR
results of \cite{Arna92a}, we found it necessary to use a box size big
enough to enclose the $2l^53l^\prime$ states in order that RR to these
states was not treated hydrogenically.  For these lowest-lying states,
the hydrogenic approximation is less valid.  Subsequent runs using a
box large enough for the $n=4$ states, and treating $n=5$ and higher
hydrogenically changed the calculated RR cross section by less than 2\%
\citep[see also the similar discussion by][]{Arna92a}.

To treat the core radiative decay in Equation \ref{path4},
where the valence electron acts as a spectator, we modify the effective
quantum number $\nu$ in the closed-channel MQDT expression by adding a
term $-i\Gamma_{core}/2$ to the core energy $E_{core}$ used in
determining $\nu$.  Here $\nu$ is a continuous variable,
calculated using $E_{cm}=E_{core}-Z^2/2\nu^2$, and
$\Gamma_{core}$ is given by
\begin{eqnarray}
\Gamma_{core}=2\pi \vert\langle 2s2p^4\vert D\vert 2s^22p^3\rangle\vert^2  \ ,
\end{eqnarray}
where $Z=19$.  We treat the valence decay in Equations
\ref{path5} and \ref{path6} hydrogenically, and add a term
$-i\Gamma_{valence}/2$ to the diagonal closed-closed part of the unphysical
scattering matrix, where
\begin{eqnarray}
\Gamma_{valence}=2\pi\sum_{n^\prime=4}^{16}\sum_{\pm 1} \vert\langle nl\vert D\vert n^\prime
(l\pm 1)\rangle\vert^2.  
\end{eqnarray}
Note that there is no interference considered between the RR pathway
in Equation \ref{path3} and the DR pathways in Equations \ref{path4},
\ref{path5}, and \ref{path6}, but this is expected to be less important
than the interference occurring between Equations \ref{path1} and \ref{path2}
since the RR rate is strongest to the lowest lying states, and only
when the RR and DR rates to the same final recombined state are
comparable will any significant interference occur.  

For \ion{F}{7}, \ion{Ar}{16}, and \ion{Fe}{25}, the present type of
R-matrix calculation has been shown to give results nearly identical
to those from the perturbative code AUTOSTRUCTURE
\citep{gorczyca96,gorczyca97a,Mitn99a}.  However, in certain
highly-sensitive cases, differences between the two codes can be seen.
For DR of \ion{Li}{2} \citep{saghiri1999}, AUTOSTRUCTURE results were
not in as good agreement with the measurements as were the R-matrix
results \citep{Pric97a}.  In \ion{Sc}{4}, AUTOSTRUCTURE calculations
needed to be extended to include interference effects between RR
and DR before agreement was found with R-matrix results
(Gorczyca et al.\ 1997).  One aim of the present work is to search for
possible interference effects in \ion{Fe}{20} where they would most
likely occur (i.e., to short-range final recombined states).  However,
for highly ionized systems, such as that studied here, the effects of
interference between the RR and DR channels are unlikely to influence
the computed Maxwellian rate coefficient (Pindzola, Badnell, \& Griffin
1992).  Indeed by comparing our AUTOSTRUCTURE calculations (which here
do not include interference effects) with our R-matrix results, we find
in the present case that these effects are negligible on the Maxwellian
rate coefficient.

R-matrix results are expected to give rise to slightly better
autoionization and/or radiative widths, compared to perturbative
approaches.  This is due to the more flexible R-matrix basis used to
describe the wavefunction of each structured continuum (i.e., continuum
with embedded resonances).  The R-matrix atomic structure calculations
start with the same $1s$, $2s$, and $2p$ orbitals and configurations as
described in \S~\ref{secauto}.  Hence the calculated level energies are
the same as for our AUTOSTRUCTURE results and prior to the final DR
cross sections calculations, these energies were shifted to the
spectroscopically known values \citep{Suga85a}.  We also calculated the
$3s$, $3p$, and $3d$ orbitals optimized on the $2s^22p^23l$
configuration-average energies.  These levels were included so that the
$2l^5 3l^\prime$ final recombined states were contained in the R-matrix
box (see discussion above).  For the resonance and continuum states all
total spin and orbital angular momenta $S^t=0-2$, $L^t=0-27$ (even and
odd parities) were used in LS-coupling, and LS-JK recoupled to include
all $J^t=0-25$ (even and odd parities).   A basis of 20 R-matrix
orbitals was used to describe each continuum $\epsilon l^\prime$ or
bound $n l^\prime$ orbital.

Using the radiation-damped R-matrix approach, the photorecombination
cross section is computed as the flux lost through the electron-ion
scattering process.  Due to the inclusion of a radiative optical
potential in the R-matrix Hamiltonian \citep{robicheaux95}, the
scattering matrix $S(E)$ is no longer unitary, and its
non-orthogonality can be related to the photorecombination cross
section as
\begin{eqnarray}
\sigma_{PR}(E)=\sum_d{\pi\over k^2}{(2J^t_d+1)\over 2(2J_{core}+1)}\sum_{\alpha}
\left\{1- \sum_{\beta}
S^*_{\alpha\beta}(E)S_{\alpha\beta}(E)\right\}\ , 
\label{eqpr}
\end{eqnarray}
where $\alpha$ is summed over all channels coupled to the initial ionic
target state $2s^22p^3(^4S_{3/2})$ and $\beta$ is summed over all open,
or continuum, channels.  The closed, or resonance, channels have been
incorporated into this scattering information via 
MQDT (Seaton 1983; Aymar, Greene, \&
Luc-Koenig 1996).  In the absence of all couplings except
for the resonance-continuum terms, Equation~\ref{eqpr} reduces to the
DR term in Equation~\ref{eqdr} plus the direct RR term and the RR/DR
interference term for those final recombined states that reside in
the box.  If all resonance states, contained in the
closed-channels, are omitted from the R-matrix expansion,
Equation~\ref{eqpr} yields just the RR cross section.  These RR results
are used for the non-resonant background to produce RR+DR results for
our AUTOSTRUCTURE, HULLAC, and MCDF results.

In order to resolve the many very narrow resonances, whose energy
positions are not known analytically, the scattering matrix $S(E)$ in
Equation~\ref{eqpr} needs to be evaluated at an enormous number of
energy points.  This is to be contrasted with the AUTOSTRUCTURE,
HULLAC, and MCDF calculations which analytically
determine the resonance energies
from a distorted wave bound state eigenvalue solution, that neglects
the accessible continua.  For the present R-matrix results, we used
800,000 points to cover the energy range $0\le E\le 120$ eV;  this gave
an energy-mesh spacing of $1.5\times 10^{-4}$ eV, which is comparable
to the $2s2p^4nl\rightarrow 2s^22p^3nl$ core radiative decay
width.  MQDT methods have been used to
minimize the computational work.  
Even with this more efficient method, however, about
{\em two days} of CPU time was required on a dual pentium pro Linux
workstation, compared to the AUTOSTRUCTURE time on the same machine of
about 40 minutes.

Our R-matrix results include the effects of radiation damping.  Despite
many of the radiative stabilizing decays here being $\Delta N=0$
transitions, using AUTOSTRUCTURE we find radiation damping to be
extremely important for \ion{Fe}{20}.  Near the Rydberg limits,
comparing the AUTOSTRUCTURE results with and without the inclusion of
the $\sum_f\Gamma^r_{df}$ term in the denominator of Equation \ref{eqdr}, we
find that there is a damping reduction by more than an order of
magnitude in the convoluted cross section.  Just as importantly,
some of the lower-$n$ resonances are damped by factors of 2 in the
convoluted cross section.  Hence, theoretical methods based on
inverse-photoionization calculations will, without the inclusion of
radiation damping, severely overestimate the true cross section,
provided that these narrow, undamped resonances are fully resolved in
the first place.

\subsection{Results}

We have multiplied the AUTOSTRUCTURE, HULLAC, and MCDF $\Delta N=0$ DR
cross sections with the relative electron-ion velocity and convolved the
results with the TSR energy spread to produce a rate coefficient for
direct comparison with our experimental results.  We have done the same
for the R-matrix RR cross section data and added the results to the
AUTOSTRUCTURE, HULLAC, and MCDF DR data.  The resulting convolved RR+DR
data are shown, respectively, in Figures~\ref{fig:FeXXresonances}(b),
\ref{fig:FeXXresonances}(c), and \ref{fig:FeXXresonances}(d).  The
R-matrix results yield a unified RR+DR cross section which we
multiplied by the relative electron-ion velocity and convolved with the
experimental energy spread.  These results are shown in
Figure~\ref{fig:FeXXresonances}(e).

Figure~\ref{fig:FeXXrates}(b) shows the AUTOSTRUCTURE, HULLAC, and MCDF
$\Delta N=0$ DR results (for $n_{max}=120$) convolved with a
Maxwell-Boltzmann electron energy distribution.  We have fitted these
DR rate coefficients using Equation~\ref{eq:drratefit}.
Table~\ref{tab:fitparameters} lists the best-fit values for the fit
parameters.  For $0.001 \le k_BT_e \le 10000$~eV, the fit is good to
better than 1.5\% for the AUTOSTRUCTURE results and 0.8\% for the MCDF
results.  The fit to the HULLAC results is good to better than 0.3\%
for $0.01 \le k_BT_e \le 10000$~eV.  Below 0.01 eV, the fit goes to
zero faster than the calculated HULLAC rate coefficient.

Because interference between the RR and DR channels appears to be
unimportant, we can also produce an R-matrix DR-only rate coefficient
($n_{max}=120$) by subtracting the RR-only R-matrix results
($n_{max}=120$) from the RR+DR results ($n_{max}=120$).  In
figure~\ref{fig:FeXXrates}(b) we show our DR-only ($n_{max}=120$) and
RR-only ($n_{max}=\infty$) results.  Table~\ref{tab:fitparameters}
lists the best-fit values for the DR fit parameters.  For $0.001 \le
k_BT_e \le 10000$~eV, the fit is good to better than 1.0\% for the
R-matrix results.  Including DR contributions from $n=120$ to $\infty$
is predicted by us to have an insignificant effect below
$k_BT_e=10$~eV, and to increase our experimentally-derived DR rate
coefficient by 1\% at 27~eV, by 3\% at 65~eV, by 5\% at 268~eV, and by
5.6\% at 10,000~eV.

Our RR rate coefficient ($n_{max}=\infty$) is listed in
Table~\ref{tab:RR}.  In order to converge at energies $\lesssim 1$~eV,
we found it necessary to top-up our R-maxtrix RR results with
hydrogenic calculations of RR into $J\ge26$ using AUTOSTRUCTURE.

\section{Discussion}
\label{sec:Discussion}

Table~\ref{tab:energylevels} gives the experimental and theoretical
energies for all \ion{Fe}{20} $n=2$ levels.  The
spectroscopically derived energies of \citet{Suga85a} are listed
first.  Also given are the unshifted energies calculated using the
AUTOSTRUCTURE, HULLAC, and MCDF techniques as well as from
calculations by \citet{Bhat89a}, \citet{Donn99a}, and \citet{Zhan00a}.
Our MCDF energies and the results of Zhang \& Pradhan agree to within
$\approx 2\%$ with the experimental values.  Our AUTOSTRUCTURE, HULLAC,
and R-matrix results and those of Bhatia et al.\ lie within $\approx 3\%$
of experiment.  The energies of Donnelly et al.\ lie within $\approx 4\%$
of the experimental values.

AUTOSTRUCTURE, HULLAC, and MCDF calculations use a perturbative
technique and yield DR resonance strengths and energies.  The R-matrix
calculations use a non-perturbative method and yield unified RR+DR
recombination results.  Comparisons of individual resonance strengths
and energies between experiment and theory are most straightforward for
perturbative calculations.  For these results the energy-integrated
resonance strength 
\begin{equation}
S_d=\int_{E_d-\Delta E/2}^{E_d+\Delta E/2} \sigma^d_{DR}(E)dE
\end{equation}
can be calculated in analytic form, thereby giving
the contribution from each isolated resonance $d$.  We compare our
experimental results with the non-perturbative R-matrix results to the
extent that is straightforwardly possible.

DR resonances are identified in Table~\ref{tab:FeXXextracteddata} by
their dominant component.  AUTOSTRUCTURE, HULLAC, and MCDF results have
been used as a guide in the resonance assignment.  In general,
unambiguous identification is possible.  One clear exception is for the
$2s 2p^4 (^4P_{3/2}) 7d_{3/2}\ (J=3)$ and $2s 2p^4 (^4P_{3/2})
7d_{5/2}\ (J=3)$ resonances.  AUTOSTRUCTURE predicts these resonances
to lie, respectively, at $\approx 0.04$ and $\approx 0.3$~eV.  MCDF
predicts them at $\approx 0.3$ and $\approx 0.04$~eV.  The ambiguity in
resonance assignment is most likely due to strong mixing between these
two states.  HULLAC predicts the $7d_{3/2}$ resonance to occur at
$\approx 0.3$~eV and that the $7d_{5/2}$ level lies below the
\ion{Fe}{19} continuum.  Our fit to the unresolved near 0~eV
recombination signal suggests this latter resonance is broad and
straddles the ionization threshold for \ion{Fe}{19}.  Whether this
level lies above or below the continuum is an example of the
uncertainty in the resonance energies typical for all calculations (see
below).

Another example of the uncertainty in the resonance energies is shown
by the unresolved near 0~eV $2s^22p^3(^2D^o_{5/2})15l$ resonance.
Our quantum defect, AUTOSTRUCTURE, and MCDF calculations find that the
$15i$ is the lowest lying DR resonance for this complex.  HULLAC
calculates that the $15f$, $g$, and $h$ levels are also DR resonances.

Figure~\ref{fig:energyratios} shows the ratio of the AUTOSTRUCTURE,
HULLAC, and MCDF resonance energies relative to the measured resonance
energies.  Below 2~eV, agreement between theory and experiment is not
that good, with discrepancies between theory and experiment of up to
30\%, 35\%, and 24\% for AUTOSTRUCTURE, HULLAC, and MCDF,
respectively.  A visual comparison between R-matrix results and
experiment finds discrepancies of up to 25\% in this energy range.  In
Figure~\ref{fig:blowup} we compare the theoretical and experimental
results.  The AUTOSTRUCTURE, MCDF, and R-matrix results, largely
predict the correct resonance strengths.  A uniform shift of the
theoretical results to lower energies would dramatically improve the
agreement between theory and experiment.  In the energy range shown,
the HULLAC results appear to be correctly predicting some of the DR
resonances and miss out on others.

An extreme example of the discrepancies of theoretical with the
measured resonance energies is shown by the resonance predicted by
AUTOSTRUCTURE, MCDF, and R-matrix (but not HULLAC) calculations to
occur at $\approx 0.04$~eV.  As discussed in
\S~\ref{sec:ExperimentalTechnique}, this resonance probably occurs at
an energy below 0.015~eV, contributing to the unresolved, near 0~eV
recombination signal.  These discrepancies of theory with experiment
below 0.8~eV makes DR of \ion{Fe}{20} an excellent case for testing
atomic structure calculations on ions with partially filled outer
shells.

For energies above 2~eV, AUTOSTRUCTURE and MCDF calculated resonance
energies agree with experiment to within 2\%.  R-matrix energies agree
with experiment to within 3\%.  HULLAC agrees with experiment to within
5\%.  The relative agreement between theory and experiment improves as
the collision energy increases.

Figure~\ref{fig:strengthratios} shows the ratio of the AUTOSTRUCTURE,
HULLAC, and MCDF resonance strengths relative to the measured resonance
strengths.  We use the data listed in
Table~\ref{tab:FeXXextracteddata}.  The mean value of this ratio is
$0.98\pm0.30(1\sigma)$ for the AUTOSTRUCTURE results,
$0.90\pm0.33(1\sigma)$ for the HULLAC results, and
$1.02\pm0.30(1\sigma)$ for the MCDF results.  These results do not
change significantly if we leave out of our analysis the weakest 10\%
of the measured resonances.  Our R-matrix results are in good agreement
with the AUTOSTRUCTURE results and show similar scatter in the
theory-to-experiment ratio of resonance strengths.  The mean values all
lie within our estimated total experimental error limits.  However, the
$1\sigma$ standard deviations for these ratios show that a significant
number of calculated resonance strengths fall outside the estimated
relative experimental uncertainty limits of $\lesssim 10\%$.

Between 0.08 and 1 eV, AUTOSTRUCTURE, HULLAC, MCDF and R-matrix
calculations all yield resonance strengths smaller than experiment.
The cause of this systematic shift is unlikely to be due to our method
for extracting resonance strengths from the experimental results.  The
spectrum between 0.08 and 1 eV is well resolved and we have a high
degree of confidence in the accuracy of the fit to the measured
non-resonant background which we subtract out to fit for the DR
resonance strengths and energies.

Shown in Figure~\ref{fig:theorystrengthratios} are the resonance
strength ratios for the AUTOSTRUCTURE/MCDF, HULLAC/MCDF, and
HULLAC/AUTOSTRUCTURE results.  The mean values of these ratios are,
respectively, $0.96\pm0.10(1\sigma)$, $0.88\pm0.26(1\sigma)$, and
$0.92\pm0.28(1\sigma)$.  These results do not change significantly if
we leave out of our analysis those resonances corresponding to the
weakest 10\% of the measured resonances.  Agreement between our
AUTOSTRUCTURE and MCDF results is good, much better than it is for
either calculation with experiment.  Our HULLAC results are in somewhat
poorer agreement with our AUTOSTRUCTURE and MCDF calculations.

A comparison between the various theoretical resonance strengths as
well as with the experimental results indicates that the HULLAC
methodology for calculating DR forming $2s^2 2p^3 nl$ resonance
configurations is incomplete.  For example, HULLAC tends to
underestimate significantly the $2s^2 2p^3 (^2D^o_{3/2,5/2}) nl$
resonance strengths and to overestimate significantly the
$2s^22p^3(^2P^o_{1/2,3/2})nl\ (l\ge3)$ resonance strengths.  These
errors are most likely due to configuration mixings induced by the
parametric potential, transferring contributions from one series to
another, and to the fact that HULLAC does not include the one-electron
operator autoionization transitions in which the initial and final
states differ by only one orbital.  These interactions can increase or
decrease the rate or have no effect at all.  Work is underway to modify
HULLAC to include the one-electron operator \citep{BarS01b}.

Another point of note is that  the AUTOSTRUCTURE and MCDF results find
a factor of $\approx 2$ drop between the resonance strength for the
$2s^22p^3(^2P^o_{1/2})21l\ (l\ge0)$ and the
$2s^22p^3(^2P^o_{1/2})22l\ (l\ge0)$ levels.  This is attributed to the
opening up of the $2s^22p^3(^2P^o_{1/2})nl \to 2s^2 2p^3(^2D_{5/2}) +
e^-$ Auger channel which reduces the radiative branching ratio by about a
half.  HULLAC results predict this Auger channel to open up between the
$2s^22p^3(^2P^o_{1/2})24l\ (l\ge0)$ and
$2s^22p^3(^2P^o_{1/2})25l\ (l\ge0)$ resonances.

There are a number of other outstanding discrepancies.  Here we only
call attention to the most glaring examples.  HULLAC underestimates the
$2s2p^4(^2P_{3/2})6d$ resonance strengths between $\approx 15-16$ eV.
HULLAC also underestimates the resonance strength for two
$2s2p^4(^2P_{3/2})6f$ resonances at 17.229 and 17.242 eV.
AUTOSTRUCTURE underestimates the $2s^2 2p^3(^2D_{3/2}^o)
17d_{3/2}\ (J=3)$ resonance strength at $\approx 0.09$~eV by a factor of
$\approx 2$.

\subsection{Rate Coefficients}

RR calculations have been carried out using R-matrix techniques and
topped up using AUTOSTRUCTURE as described above.  \citet{Arna92a}
have calculated the rate coefficient for RR of \ion{Fe}{20} and
presented a fit to their results which is supposed to be valid between
10$^5$ and 10$^8$ K.  Their results are plotted in
Figure~\ref{fig:FeXXrates}(a).  We find that their rate coefficient
agrees with our R-matrix results to within 10\% for $k_BT_e$ of between
$\approx 10^{3.4}$ and $\approx 10^{7.8}$ K.

The calculations of \citet{Jaco77a} and Roszman
\citep{Arna92a} were carried out using perturbative techniques,
but they only published Maxwellian-averaged rate
coefficients.  \citet{Savi99a} demonstrated that comparisons of
only Maxwellian-averaged rate coefficients 
cannot be used reliably to distinguish
between different theoretical techniques.  Disagreement between
experiment and theory can be used to demonstrate the inadequacy of
a particular theoretical technique.  However, agreement between experiment
and theory can be fortuitous.  A detailed comparison of resonance
strengths and energies is the only way to verify the accuracy of
DR rate coefficient calculations.  Unfortunately, neither Jacobs et al.\
nor Roszman published their calculated resonance strengths and energies.

Figure~\ref{fig:FeXXrates}(a) shows the theoretical $\Delta N=0$ DR
rate coefficients of Jacobs et al.\ as fitted by \citet{Shul82a} and of
Roszman as reported by \citet{Arna92a}.  \ion{Fe}{20} is predicted to
peak in fractional abundance in an optically thin, low-density
photoionized plasma of cosmic abundances at $k_BT_e \approx 35$~eV
\citep{Kall01a}.  At this temperature, our experimentally derived DR
rate coefficient is a factor of $\approx 1.8$ larger than the rate
coefficient of Roszman and of $\approx 4$ times larger than the rate
coefficient of Jacobs et al.\  The reason for these disrepancies is most
likely because these calculations did not correctly predict the DR
resonance structure at the relevant energies.  Also, neither
calculation accounts for DR via $2p_{1/2} \to 2p_{3/2}$ core
excitations.  The experimentally-derived DR rate coefficient is
$\approx 4$ times larger than the RR rate coefficient at $k_BT_e
\approx 35$~eV.

We have calculated the $\Delta N=0$ rate coefficient for DR of
\ion{Fe}{20} using our AUTOSTRUCTURE, HULLAC, MCDF, and R-Matrix
techniques.  The results are shown in Figure~\ref{fig:FeXXrates}(b).
For $k_BT_e \gtrsim 10$~eV, our experimental and theoretical results agree
to better than $\approx 15\%$.  This temperature range includes the
predicted zone of formation for \ion{Fe}{20} in a photoionized plasma
of cosmic abundances.  We note that for $k_BT_e \ge 100$~eV, $N=2 \to
N^\prime=3$ DR begins to contribute more than 10\% to the total DR rate
coefficient \citep{Arna92a}.  We plan to measure DR via this core
excitation at a future date.  Agreement below $k_BT_e \lesssim 1$~eV is
difficult to quantify due to current theoretical and experimental
limitation for studying resonances near 0~eV.

\section{Conclusions}
\label{sec:Conclusions}

We have measured the resonance strengths and energies for $\Delta N=0$
DR of \ion{Fe}{20}.  The relative experimental uncertainty is
estimated at $\lesssim 10\%$ and the total experimental uncertainty at
$\lesssim 20\%$.  We have also calculated resonance strengths and
energies using the state-of-the art AUTOSTRUCTURE, HULLAC, MCDF, and
R-matrix methods.  On average we find good agreement between the
theoretical and experimental resonance strengths.  However, a large
number of the theoretical resonance strengths differ from the measured
values by more than three times the relative experimental uncertainty
limits.  These discrepancies suggest errors in the calculated level
populations and line emission spectrum for the recombined ions.

We have used our experimental and theoretical results to produce
Maxwellian-averaged rate coefficients for $\Delta N=0$ DR of
\ion{Fe}{20}.  For $k_BT_e \gtrsim 10$~eV (which includes the predicted
temperature of formation for \ion{Fe}{20} in a photoionized plasma),
theory and experiment agree to better than $\approx 15\%$.  Apparently
many of the discrepancies between the theoretical and experimental
resonance strengths average away when one calculates the
Maxwellian-averaged rate coefficient.

Agreement for $k_BT_e \lesssim 1$~eV is difficult to quantify due to
current theoretical and experimental limitation.  Published
$LS$-coupling DR rate coefficients are in poor agreement with
experiment for $k_BT_e \lesssim 80$~eV.  Lastly, we have calculated the
rate coefficient for RR of \ion{Fe}{20}.  Our RR results are in good
agreement with published calculations.

\acknowledgements

This work was supported in part by NASA High Energy Astrophysics X-Ray
Astronomy Research and Analysis grant NAG5-5123 and NASA Space
Astrophysics Research and Analysis Program grant NAG5-5261.  Travel to
and living expenses at TSR for DWS were funded by NATO Collaborative
Research Grant CRG-950911.  The experimental work has been supported in
part by the German Federal Minister for Education and Research (BMBF)
under Contract Nos.\ 06 GI 475, 06 GI 848, and 06 HD 854I.  NRB was
supported in part by PPARC through a grant (PPA/G/S/1997/00783) to the
University of Strathclyde.  Work performed at Lawrence Livermore
National Laboratory was under the auspices of the US Department of
Energy by the University of California, Lawrence Livermore National
Laboratory, under Contract number W-7405-ENG-48.  TWG was supported in
part by the NSF through a grant to the Institute for Theoretical Atomic
and Molecular Physics at Harvard University and the Smithsonian
Astrophysical Observatory.

\vfill
\clearpage
\eject

\begin{deluxetable}{lrrrrrrr}
\tabletypesize{\scriptsize}
\footnotesize
\rotate
\tablecaption{Experimental and (unshifted) theoretical energy levels
(relative to the ground state) for the $n=2$ shell of \protect\ion{Fe}{20}.
\label{tab:energylevels}}
\tablewidth{0pt}
\tablehead{\multicolumn{1}{c}{Level} & \multicolumn{7}{c}{Energy (eV)}\\
\cline{2-8}
\colhead{ } & \colhead{Experiment\tablenotemark{a}}
& \colhead{AUTOSTRUCTURE\tablenotemark{b,c}}
& \colhead{Bahtia et al.\tablenotemark{d}}
& \colhead{Donnelly et al.\tablenotemark{e}}
& \colhead{HULLAC\tablenotemark{b}}
& \colhead{MCDF\tablenotemark{b}}
& \colhead{Zhang \& Pradhan\tablenotemark{f}}
\\
}
\startdata
$2s^2 2p^3\ ^4S^o_{3/2} $ &       0 &       0 &       0 &       0 &       0
&        0 &        0\\
$2s^2 2p^3\ ^2D^o_{3/2} $ & 17.1867 & 17.4428 & 17.5337 & 17.5652 & 17.3514
& 17.4848 & 17.400\\
$2s^2 2p^3\ ^2D^o_{5/2} $ & 21.8373 & 22.5298 & 22.5013 & 22.6023 & 22.4259
& 22.2628 & 21.376\\
$2s^2 2p^3\ ^2P^o_{1/2} $ & 32.2694 & 32.1386 & 32.3219 & 32.49241& 31.9788
& 32.1682 & 32.245\\
$2s^2 2p^3\ ^2P^o_{3/2} $ & 40.0890 & 39.9930 & 40.2327 & 40.29434& 40.0720
& 40.0987 & 39.908\\
$2s^{\ } 2p^4\ ^4P_{5/2}$ & 93.3266 & 92.9521 & 93.1364 & 93.43567& 93.4074
& 93.2280 & 93.198\\
$2s^{\ } 2p^4\ ^4P_{3/2}$ & 101.769 &101.1239 & 101.429 & 101.5764&101.5300
& 101.906 & 101.30\\
$2s^{\ } 2p^4\ ^4P_{1/2}$ & 104.486 &103.8390 & 104.154 & 104.2588&104.3240
& 104.592 & 103.99\\
\enddata
\tablenotetext{a}{\citet{Suga85a}.}
\tablenotetext{b}{Present results.}
\tablenotetext{c}{Also for R-matrix results (see \S~\ref{secrmat}.)}
\tablenotetext{d}{\citet{Bhat89a}.}
\tablenotetext{e}{\citet{Donn99a}.}
\tablenotetext{f}{\citet{Zhan00a}.}
\end{deluxetable}

\vfill
\clearpage
\eject

\begin{deluxetable}{lccccccccc}
\tabletypesize{\scriptsize}
\footnotesize
\rotate
\tablecaption{Comparison of the measured and calculated resonance
energies $E_d$ and energy-integrated cross sections $S_d$
for \protect\ion{Fe}{20} to \protect\ion{Fe}{19} $\Delta N=0$ DR.
\label{tab:FeXXextracteddata}}
\tablewidth{0pt}
\tablehead{
\colhead{ } & \multicolumn{4}{c}{$E_d$ (eV)} & \colhead{ } &
\multicolumn{4}{c}{$S_d$ ($10^{-21}$ cm$^2$ eV)} \\ %$\hat{\sigma}_i$
\cline{2-5} \cline{7-10} \\
\colhead{Dominant component} & \colhead{AUTOSTRUCTURE \tablenotemark{a}} &
\colhead{HULLAC\tablenotemark{a}} & \colhead{MCDF\tablenotemark{a}} &
\colhead{Experiment\tablenotemark{b,c}}  &
\colhead{ } & \colhead{AUTOSTRUCTURE} & \colhead{HULLAC} & \colhead{MCDF} &
\colhead{Experiment\tablenotemark{b}}
}
\startdata
$2s^22p^3(^2D_{5/2})15f$\tablenotemark{d,e}
& & 6.407E-02 & & &
& & 11435.9 & & \\
$2s^22p^3(^2D_{5/2})15g$\tablenotemark{d,e}
& & 8.774E-02 & & &
& & 3660.8 & & \\
$2s^22p^3(^2D_{5/2})15h$\tablenotemark{d,e}
& & 9.276E-02 & & &
& & 1212.3 & & \\
$2s^22p^3(^2D_{5/2})15i$\tablenotemark{d}
& 4.734E-03 & 9.604E-02 & 4.0E-04 & 1.163E-03 &
& 17991.0 & 373.1 & 85381 & 141882.0 \\
$2s^22p^3(^2D_{5/2})15k$\tablenotemark{d}
& 5.216E-03 & 9.883E-02 & 8.0E-04 & 2.099E-03 &
& 7494.9 & $\ll0.1$ & 51801 & 105417.0 \\
$2s^22p^3(^2D_{5/2})15l$\tablenotemark{d}
& 5.591E-03 & 0.1008 & 1.0E-03 & 2.801E-03 &
& 3633.4 & $\ll0.1$ & 13740 &  63979.5 \\
$2s^22p^3(^2D_{5/2})15m$\tablenotemark{d}
& 5.893E-03 & 0.1008 & 1.4E-03 & 3.347E-03 &
& 2072.4 & $\ll0.1$ & 6064 &  25492.4 \\
$2s^22p^3(^2D_{5/2})15n$\tablenotemark{d}
& 6.140E-03 & 0.1008 & 1.8E-03 & 3.783E-03 &
& 1270.4 & $\ll0.1$ & 3234 &  10574.9 \\
\enddata

\tablenotetext{a}{Weighted energy:
$E_d=\sum E_d S_d / \sum S_d$.}
\tablenotetext{b}{1$\sigma$ statistical fitting uncertainties only.}
\tablenotetext{c}{Absolute energy scale uncertainty $\lesssim 0.5\%$ for
$E \lesssim 25$~eV and $\lesssim 0.2\%$ for $E \gtrsim 25$~eV.}
\tablenotetext{d}{See \S~\ref{sec:Results} about the experimental results.}
\tablenotetext{e}{See \S~\ref{sec:Discussion} about the theoretical results.}
\tablenotetext{f}{Unable to fit for resonance energy.}

\end{deluxetable}

\begin{deluxetable}{cccccc}
\rotate
\tablecaption{Rate coefficient fit parameters for $\Delta N=0$ DR of
\protect\ion{Fe}{20} forming \protect\ion{Fe}{19} ($n_{max}=120$).  The
units are cm$^3$~s$^{-1}$~K$^{1.5}$  for $c_i$ and eV for $E_i$.
\label{tab:fitparameters}}
\tablewidth{0pt}
\tablehead{\colhead{Parameter} & \colhead{Experiment} & \colhead{AUTOSTRUCTURE}
& \colhead{HULLAC} & \colhead{MCDF} & \colhead{R-matrix}}
\startdata
c$_1$	& 8.53E-05 & 1.38E-05 & 6.25E-05 & 4.51E-06 & 9.80E-06 \\
c$_2$	& 1.62E-04 & 3.29E-05 & 9.68E-05 & 2.88E-06 & 3.47E-05 \\
c$_3$	& 7.71E-05 & 3.02E-06 & 2.69E-04 & 1.41E-04 & 3.25E-05 \\
c$_4$	& 4.04E-05 & 1.96E-04 & 3.19E-04 & 8.29E-05 & 3.04E-05 \\
c$_5$	& 1.75E-04 & 5.17E-04 & 3.08E-03 & 1.84E-04 & 1.35E-04 \\
c$_6$	& 6.73E-04 & 2.49E-03 & 3.01E-03 & 5.45E-04 & 1.30E-04 \\
c$_7$	& 2.85E-03 & 2.92E-03 & 7.62E-03 & 3.37E-03 & 6.57E-04 \\
c$_8$	& 3.18E-03 & 2.13E-03 & 1.55E-02 & 3.98E-03 & 3.06E-03 \\
\enddata
\end{deluxetable}

\begin{deluxetable}{lclclclc}
\tablecaption{R-matrix rate coefficient for RR of \protect\ion{Fe}{20}
forming \protect\ion{Fe}{19} as a function of plasma temperature for
$n_{max}=\infty$.  The R-matrix results have been topped-up using
AUTOSTRUCTURE for RR into $J \ge 26$ levels.
\label{tab:RR}}
\tablewidth{0pt}
\tablehead{
\colhead{$k_BT_e$} & \colhead{Rate} & \colhead{$k_BT_e$} & \colhead{Rate} &
\colhead{$k_BT_e$} & \colhead{Rate} & \colhead{$k_BT_e$} & \colhead{Rate}\\
\colhead{(eV)} & \colhead{(cm$^3$~s$^{-1}$)} & \colhead{(eV)} & \colhead{(cm$^3$~s$^{-1}$)} &\colhead{(eV)} & \colhead{(cm$^3$~s$^{-1}$)} & \colhead{(eV)} & \colhead{(cm$^3$~s$^{-1}$)}}
\startdata
0.001 & 1.347E-08 & 0.1 & 1.082E-09 & 10  & 6.071E-11 & 1000  & 1.729E-12 \\
0.002 & 9.578E-09 & 0.2 & 7.277E-10 & 20  & 3.832E-11 & 2000  & 8.811E-13 \\
0.003 & 7.759E-09 & 0.3 & 5.706E-10 & 30  & 2.896E-11 & 3000  & 5.814E-13 \\
0.004 & 6.639E-09 & 0.4 & 4.777E-10 & 40  & 2.353E-11 & 4000  & 4.286E-13 \\
0.005 & 5.871E-09 & 0.5 & 4.154E-10 & 50  & 1.995E-11 & 5000  & 3.364E-13 \\
0.006 & 5.309E-09 & 0.6 & 3.699E-10 & 60  & 1.740E-11 & 6000  & 2.749E-13 \\
0.007 & 4.875E-09 & 0.7 & 3.354E-10 & 70  & 1.549E-11 & 7000  & 2.311E-13 \\
0.008 & 4.530E-09 & 0.8 & 3.082E-10 & 80  & 1.400E-11 & 8000  & 1.984E-13 \\
\enddata
\end{deluxetable}

\begin{figure}
\plotone{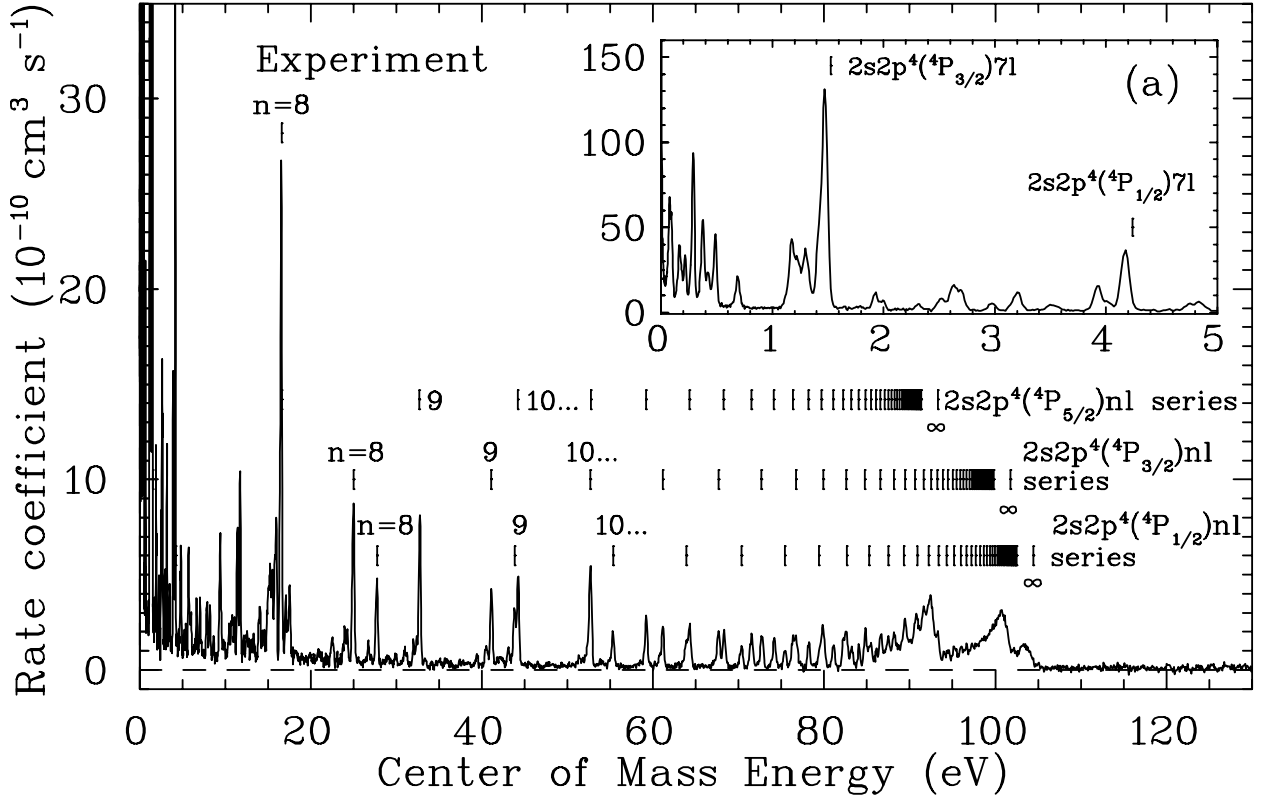}
\caption{\protect\ion{Fe}{20} to \protect\ion{Fe}{19} $\Delta N=0$ DR
resonance structure: (a) Experimental, (b) AUTOSTRUCTURE, (c) HULLAC,
(d) MCDF, and (e) R-matrix results.  The experimental and theoretical
data represent the DR and RR cross sections times the electron-ion
relative velocity convolved with the energy spread of the experiment
(i.e., a rate coefficient) and are shown versus electron-ion
center-of-mass collision energy.  In (a) resonances resulting from the
$^4S^o_{3/2}\ -\ ^4P_{5/2}$, $^4S^o_{3/2}\ -\ ^4P_{3/2}$, and
$^4S^o_{3/2}\ -\ ^4P_{1/2}$ core excitations are labeled for capture
into high $l$ levels.  Unlabeled resonances are due to capture into low
$l$ levels or due to DR via other core excitations.  Many of the
unlabeled resonances below $\approx 40$~eV are due to DR via
$2p_{1/2}-2p_{3/2}$ core excitations.  The nonresonant ``background''
rate coefficient in (a) is due primarily to RR.  In (b), (c), and (d)
we have added the convolved, non-resonant RR contribution obtained from
our R-matrix calculations to our DR results.}
\label{fig:FeXXresonances}
\end{figure}

\setcounter{figure}{0}
\begin{figure}
\plotone{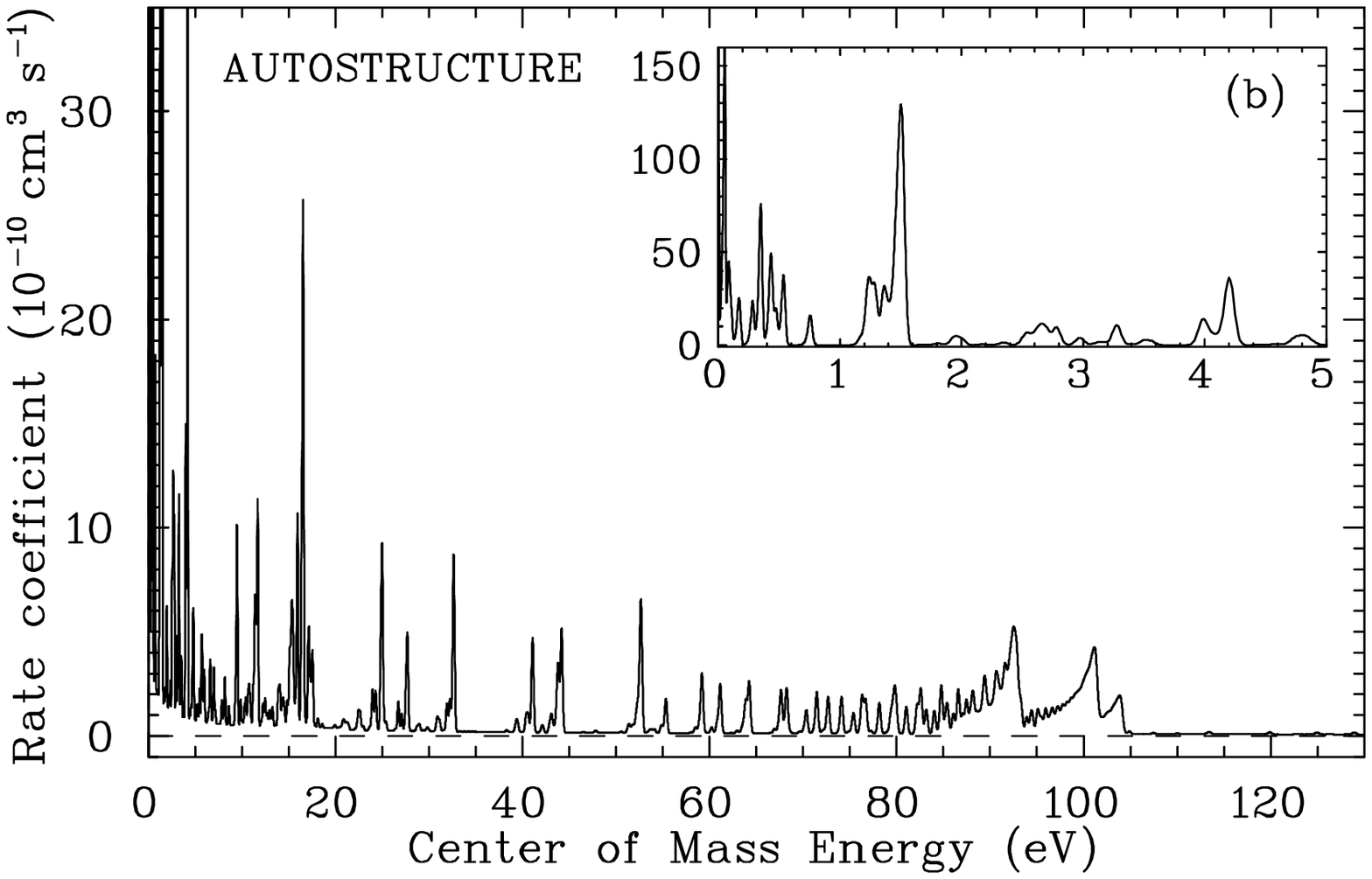}
\plotone{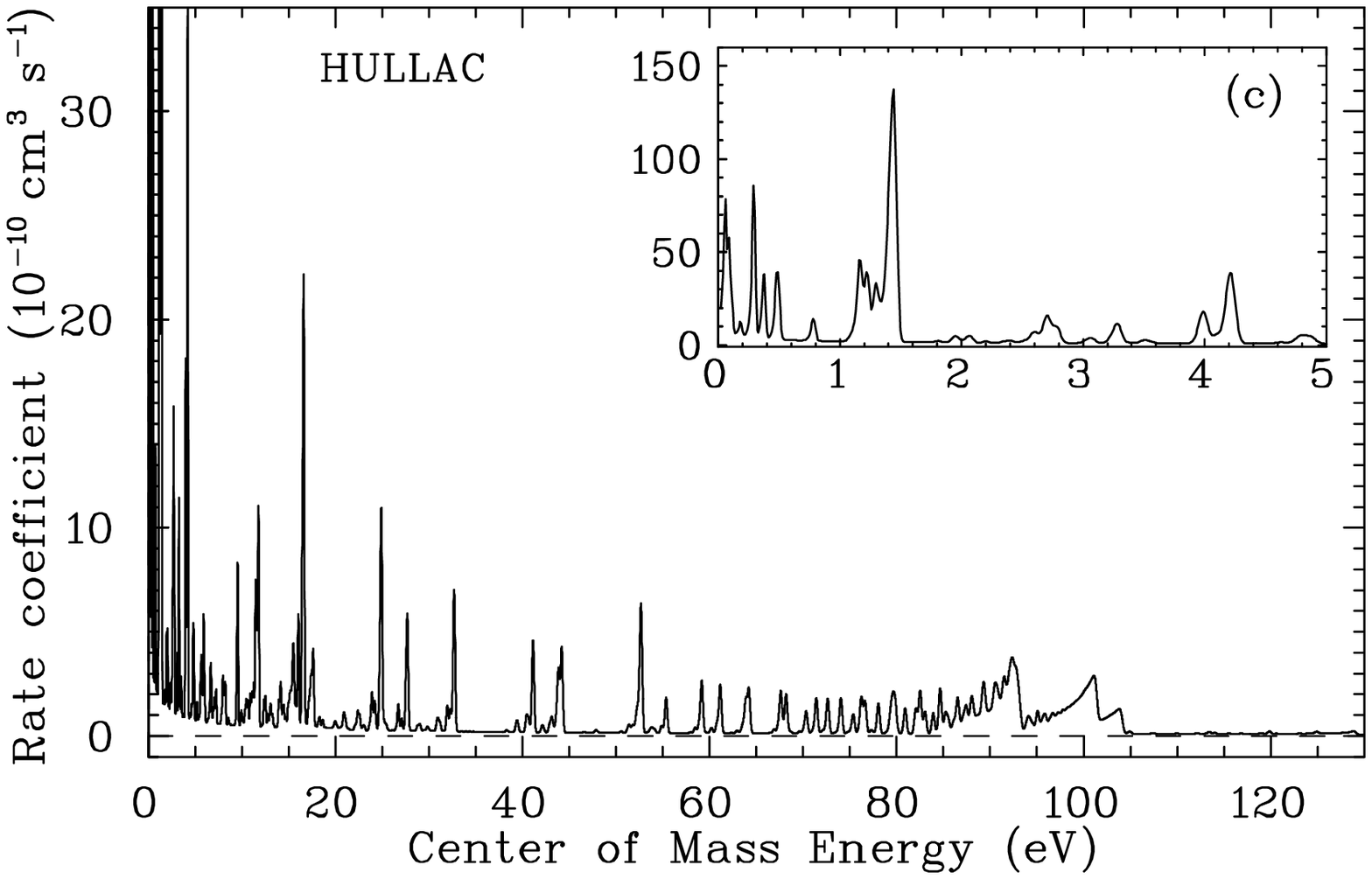}
\caption{\it Continued}
\end{figure}

\setcounter{figure}{0}
\begin{figure}
\plotone{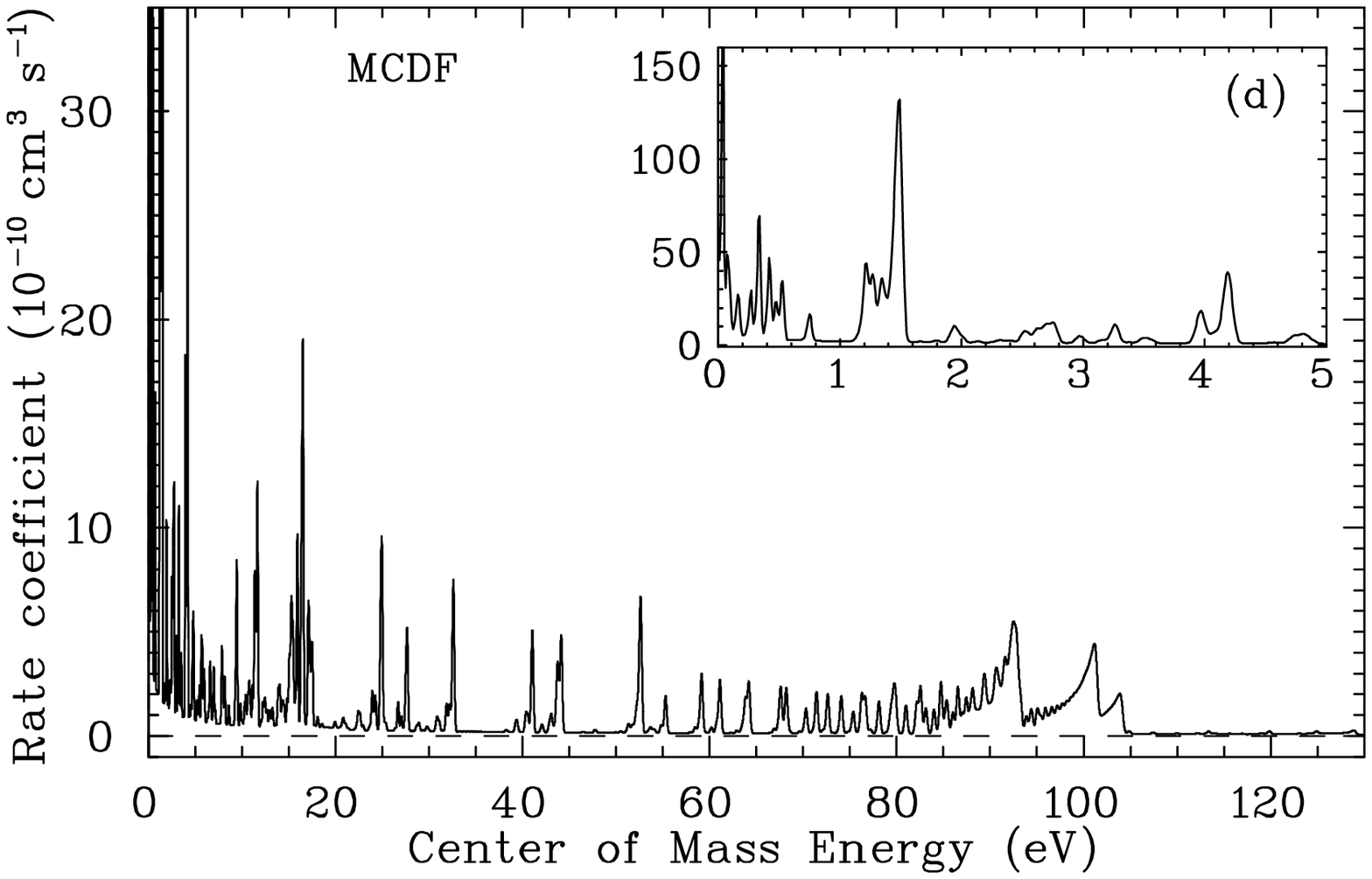}
\plotone{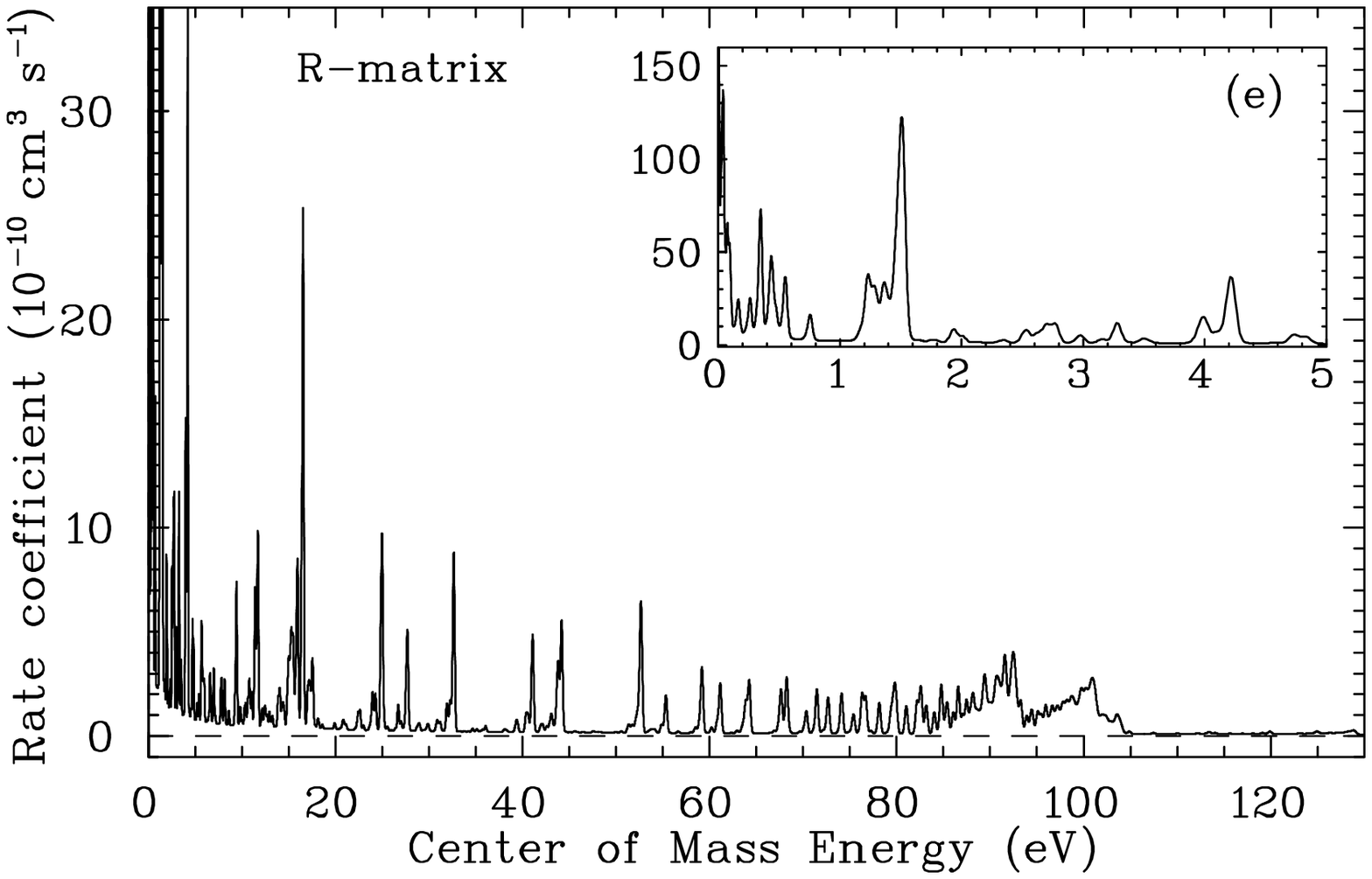}
\caption{\it Continued}
\end{figure}

\setcounter{figure}{1}
\begin{figure}
\plotone{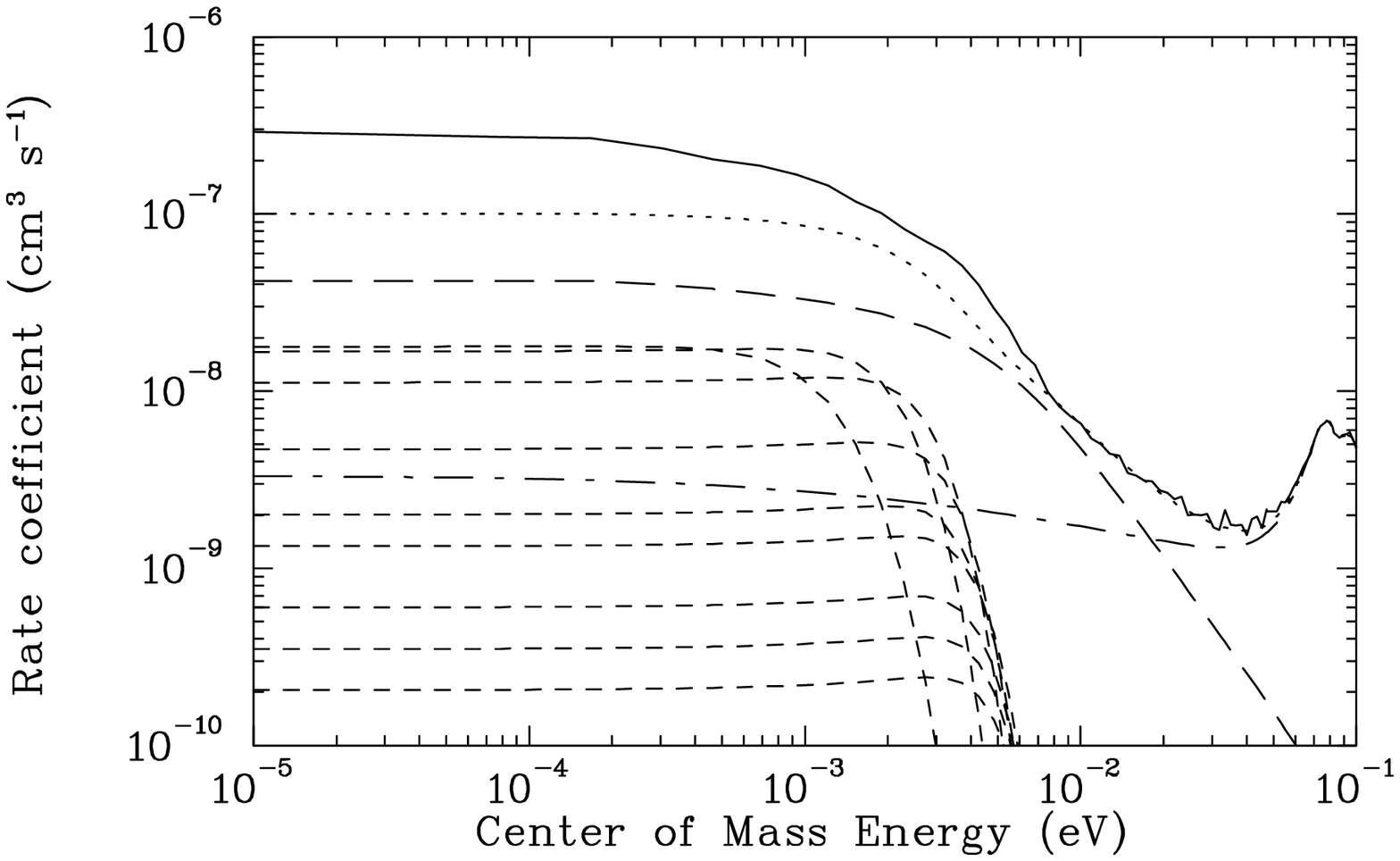}
\caption{Measured and fitted \protect\ion{Fe}{20} to
\protect\ion{Fe}{19} $\Delta N=0$ DR resonance structure below 0.1~eV.
The experimental results are shown by the {\it solid curve}.  The {\it
dotted-long-dashed curve} is the fit to the data using our calculated
RR rate coefficient and taking into account all resolved resonances.
The {\it dotted curve} is the fit including the estimated contributions
from the unresolved $2s^22p^3(^2D^o_{5/2})15l$ ({\it short-dashed
curves}) and $2s 2p^4 (^4P_{3/2}) 7d$ ({\it long dashed curve})
resonances (see \S~\ref{sec:Results}).}
\label{fig:nearzero}
\end{figure}

\begin{figure}
\plotone{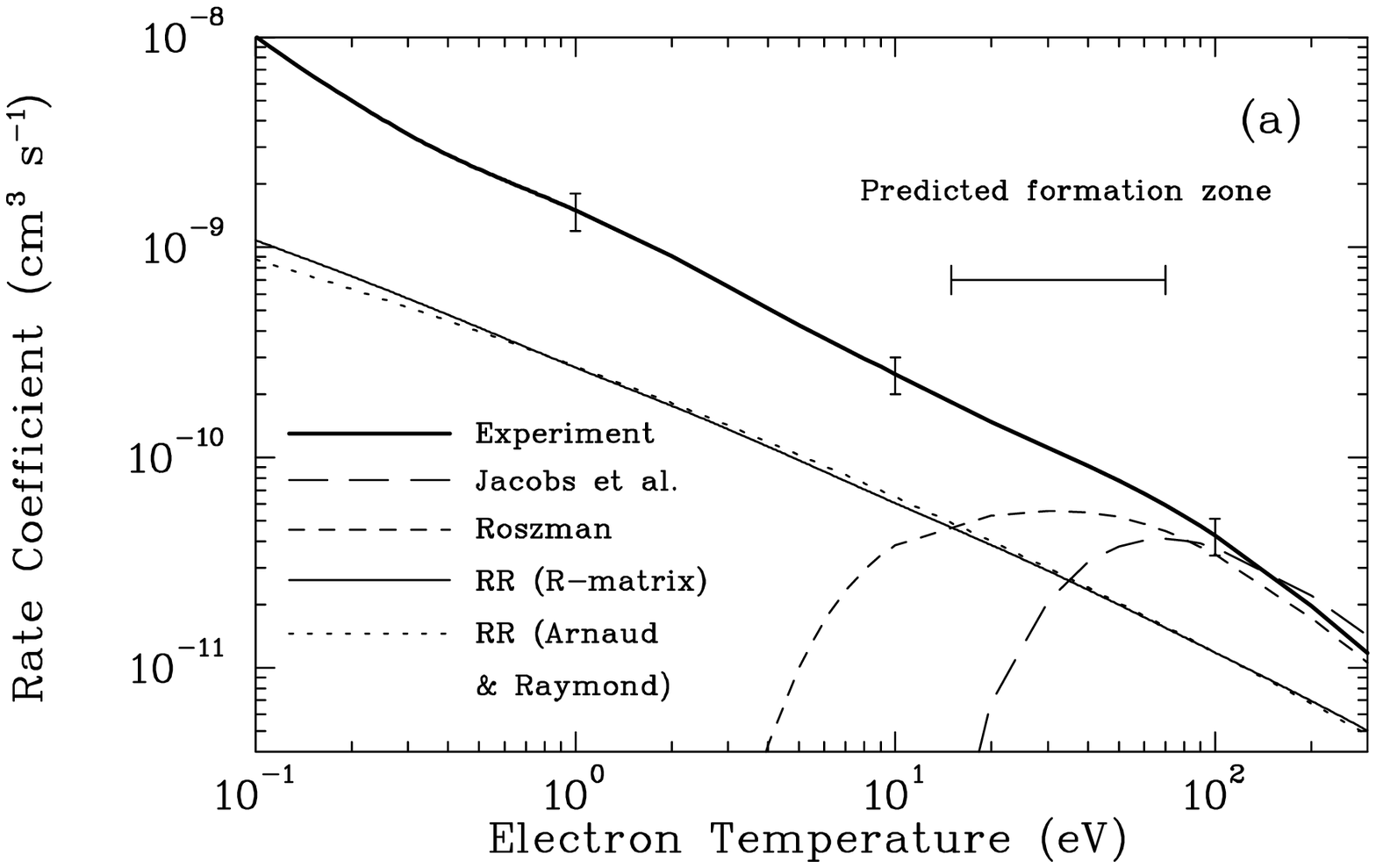}
\caption{\protect\ion{Fe}{20} to \protect\ion{Fe}{19}
Maxwellian-averaged rate coefficient for $\Delta N=0$ DR from
$k_BT_e=0.1$ to 300~eV.  (a) The {\it thick solid curve} represents our
experimentally-derived rate coefficient using the results shown in
Figure~\ref{fig:FeXXresonances}(a) and listed in
Table~\ref{tab:FeXXextracteddata}.  The error bars represent the
estimated maximum experimental uncertainty of 20\% for $k_BT_e \ge
10$~eV.  The {\it long-dashed curve} shows the $LS$-coupling
calculations of \protect\citet{Jaco77a} as fitted by
\protect\citet{Shul82a}.  The {\it short-dashed curve} shows the
unpublished $LS$-coupling calculations of Roszman as given by
\protect\citet{Arna92a}.  The {\it thin solid curve} is our R-matrix RR
rate coefficient ($n_{max}=\infty$) which has been topped up using
AUTOSTRUCTURE.  Also shown is the recommended RR rate coefficient of
Arnaud \& Raymond (1992; {\it dotted curve}).}
\label{fig:FeXXrates}
\end{figure}

\setcounter{figure}{2}
\begin{figure}
\plotone{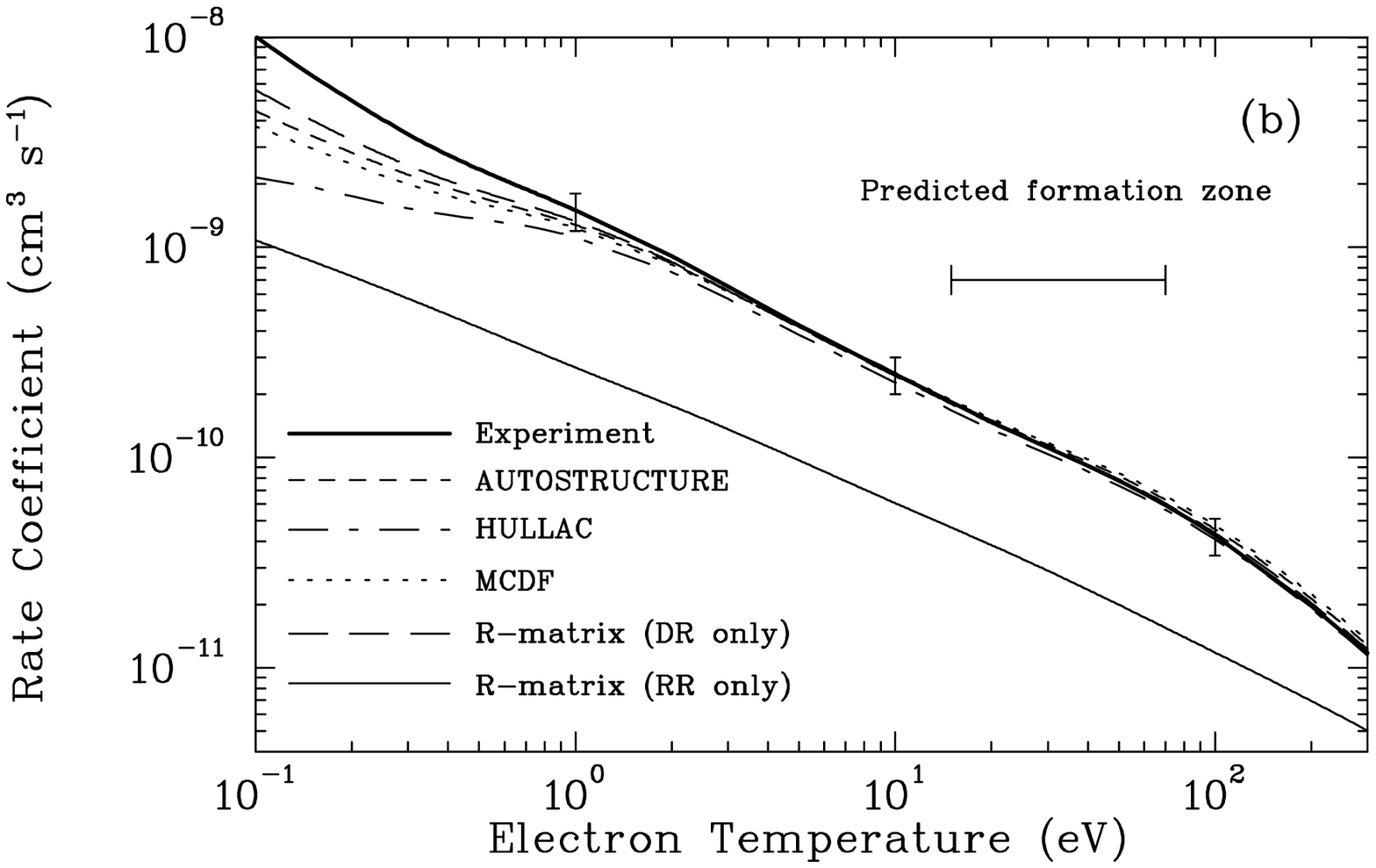}
\caption{{\it Continued.} (b) In addition to our experimentally-derived
DR rate coefficient ({\it thick solid curve}) and our topped up
R-matrix RR rate coefficient ({\it thin solid curve}), both from (a),
we also show our AUTOSTRUCTURE ({\it short-dashed curve}), HULLAC ({\it
dotted-long-dashed curve}), MCDF ({\it dotted curve}), and R-matrix
results (minus the R-matrix RR contribution, {\it long dashed curve}).
All DR rate coefficients in (b) are for an $n_{max}=120$.  The
formation zone for \protect\ion{Fe}{20} for an optically thin, low-density
photoionized plasma of cosmic abundances as predicted by XSTAR
\protect\citep{Kall01a} is shown by the horizontal solid line in both (a) 
and (b).}
\end{figure}

\begin{figure}
\plotone{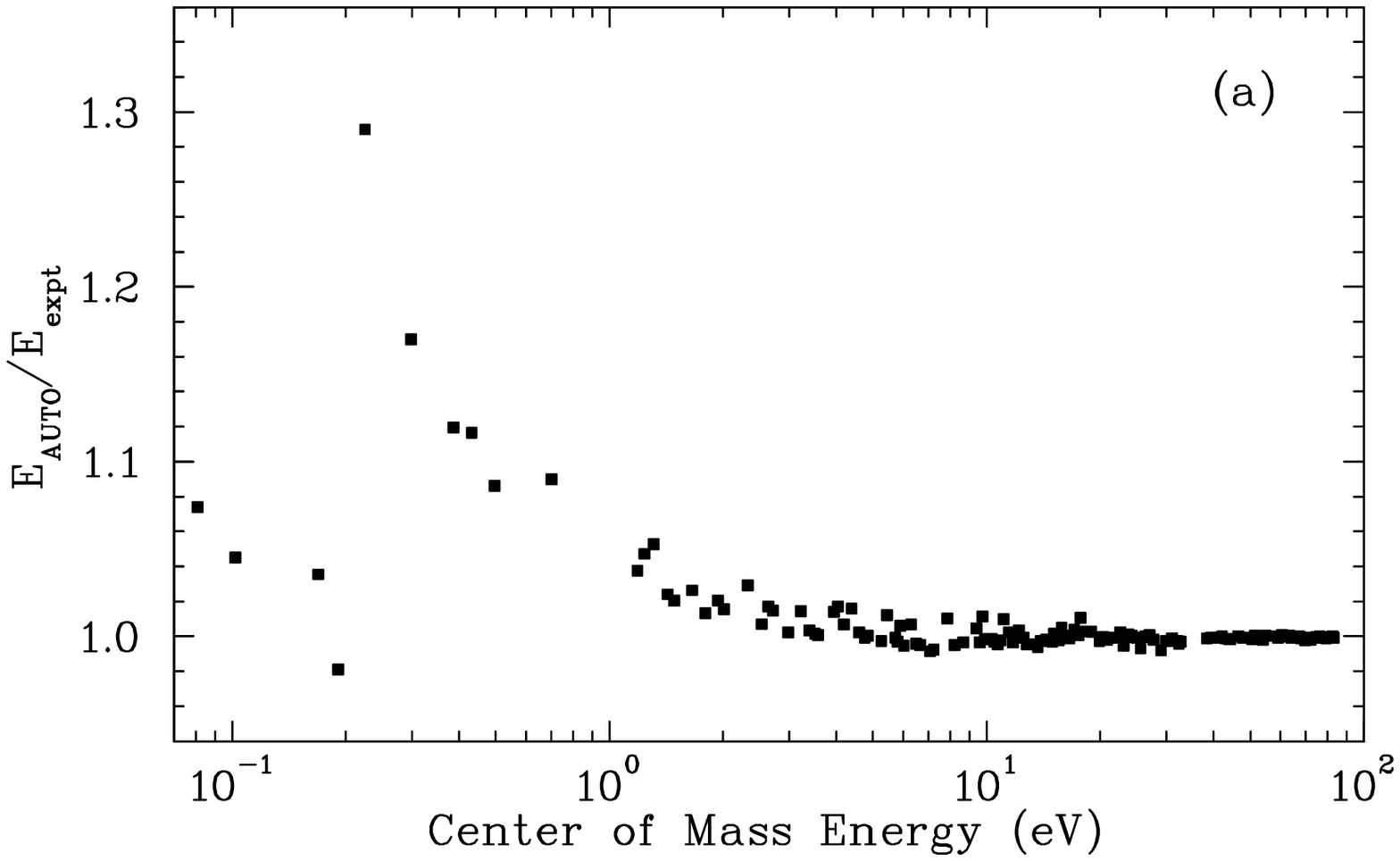}
\caption{The ratio of the (a) AUTOSTRUCTURE, (b) HULLAC, and (c) MCDF
resonance energies relative to the measured resonance energies as a
function of center-of-mass collision energy from 0.07 to 100 eV.}
\label{fig:energyratios}
\end{figure}

\setcounter{figure}{3}
\begin{figure}
\plotone{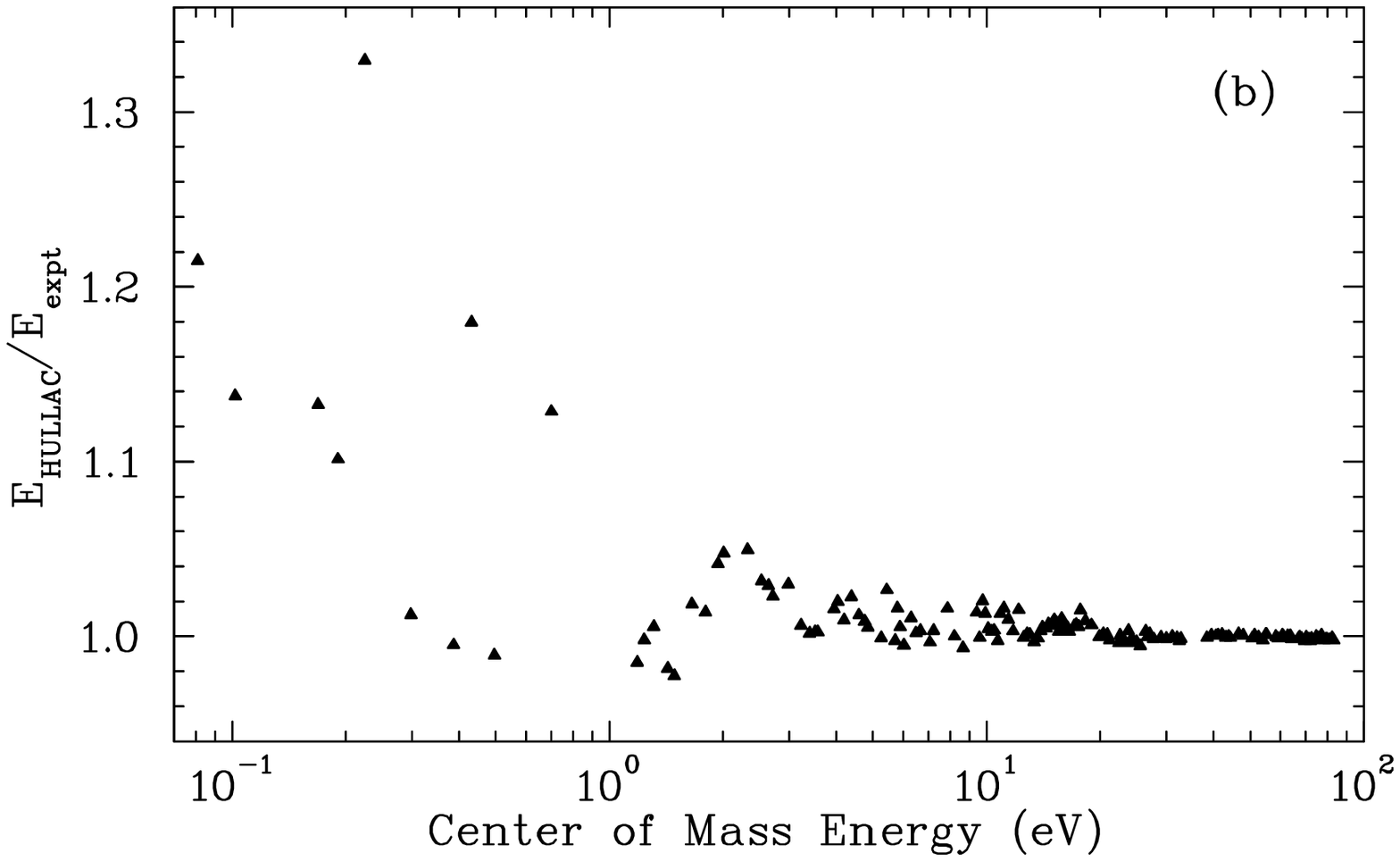}
\plotone{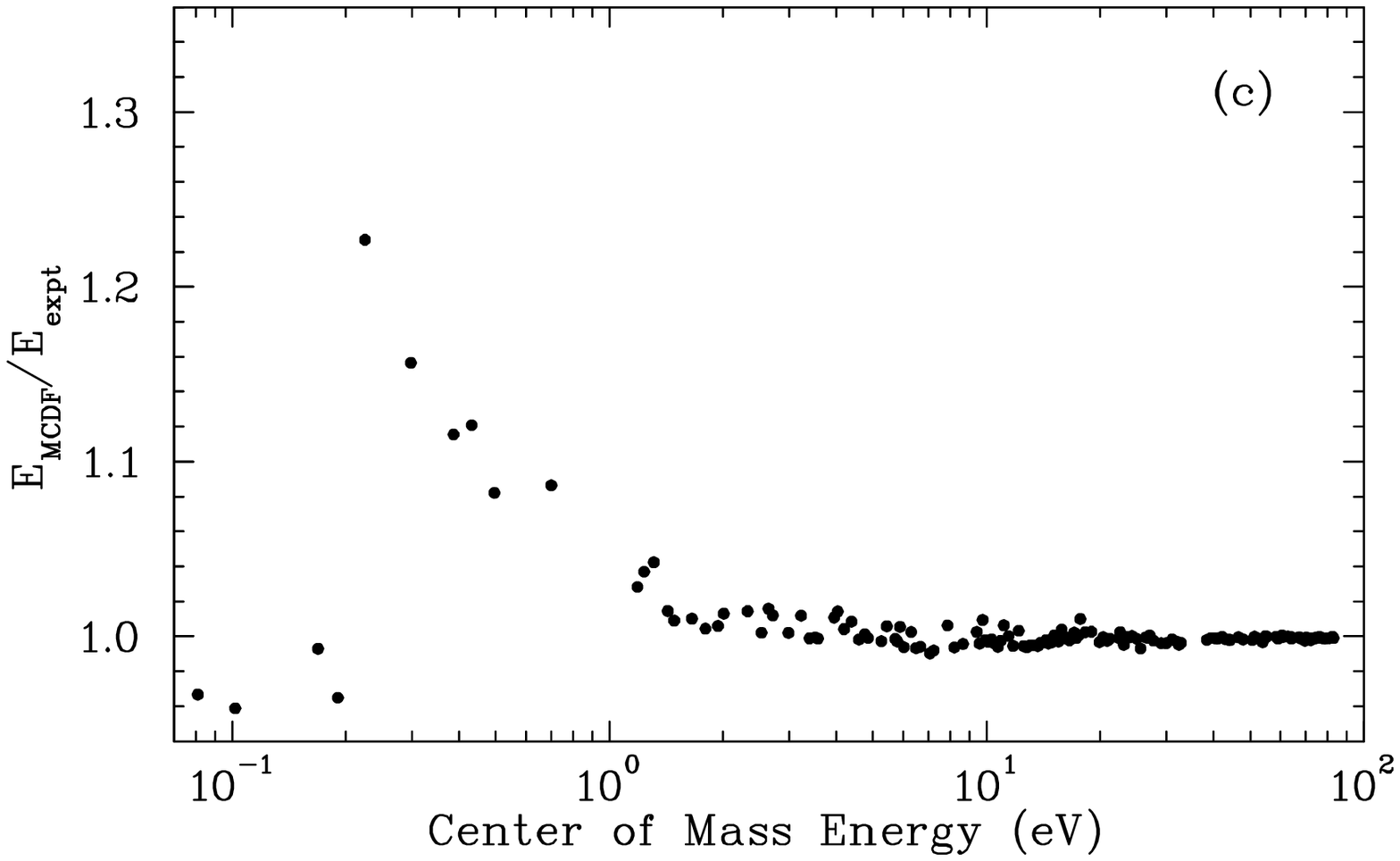}
\caption{\it Continued}
\end{figure}

\begin{figure}
\plotone{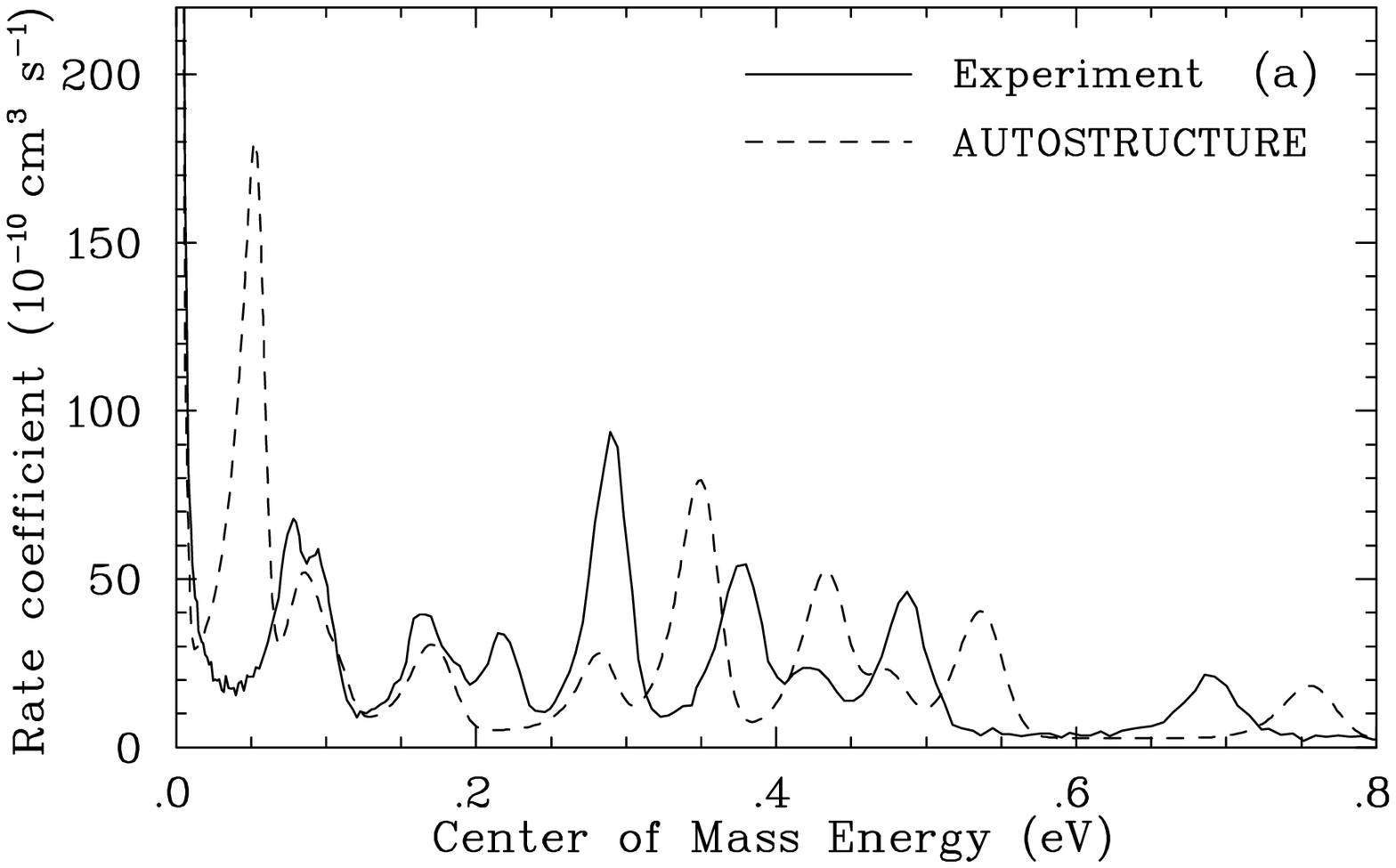}
\caption{Theoretical \protect\ion{Fe}{20} to \protect\ion{Fe}{19}
$\Delta N=0$ DR resonance structure between 0.015 and 0.8 eV compared
to our experimental results: (a) AUTOSTRUCTURE, (b) HULLAC, (c) MCDF,
and (d) R-matrix results.  See Figure~\ref{fig:FeXXresonances} for
details.}
\label{fig:blowup}
\end{figure}

\setcounter{figure}{4}
\begin{figure}
\plotone{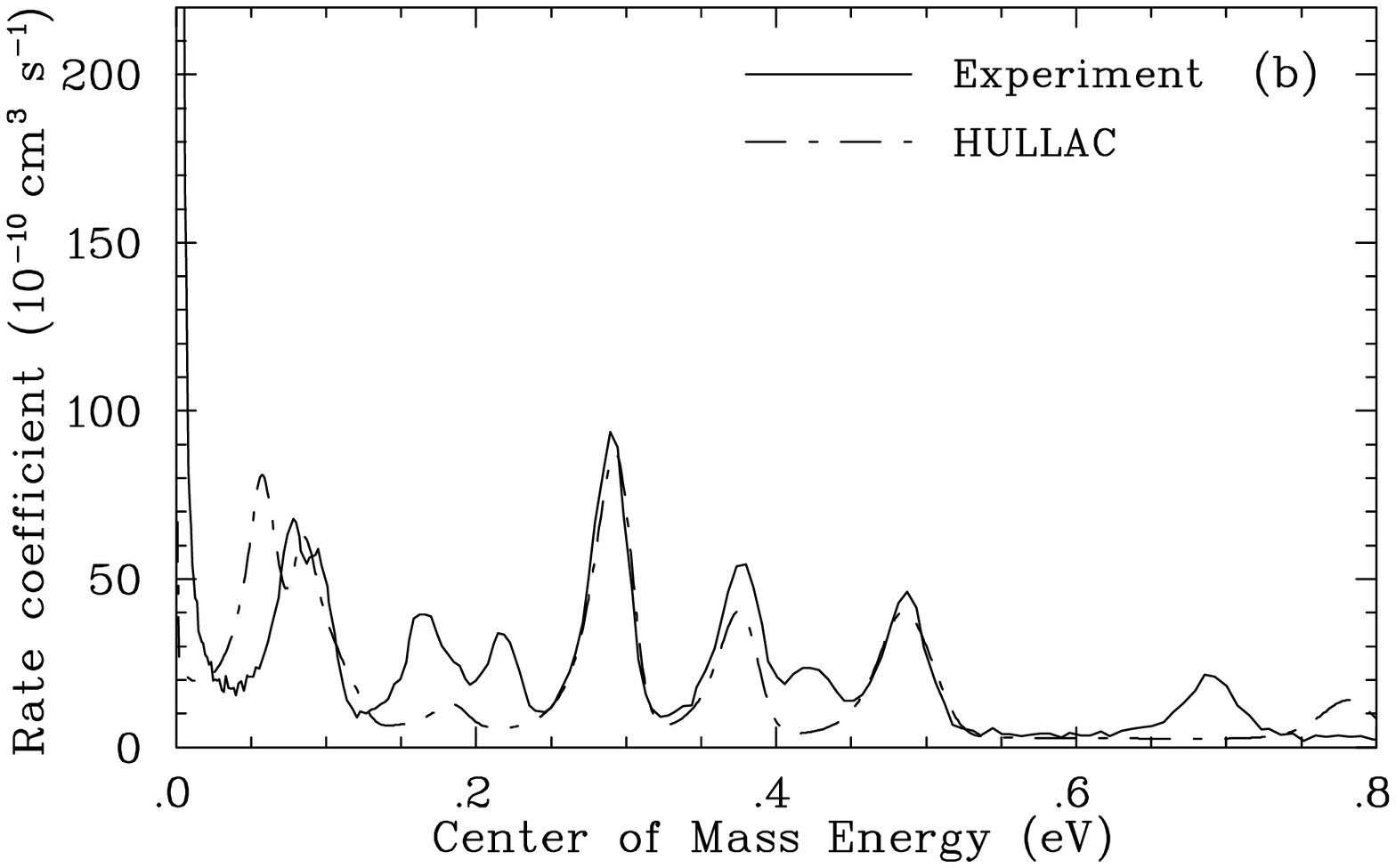}
\plotone{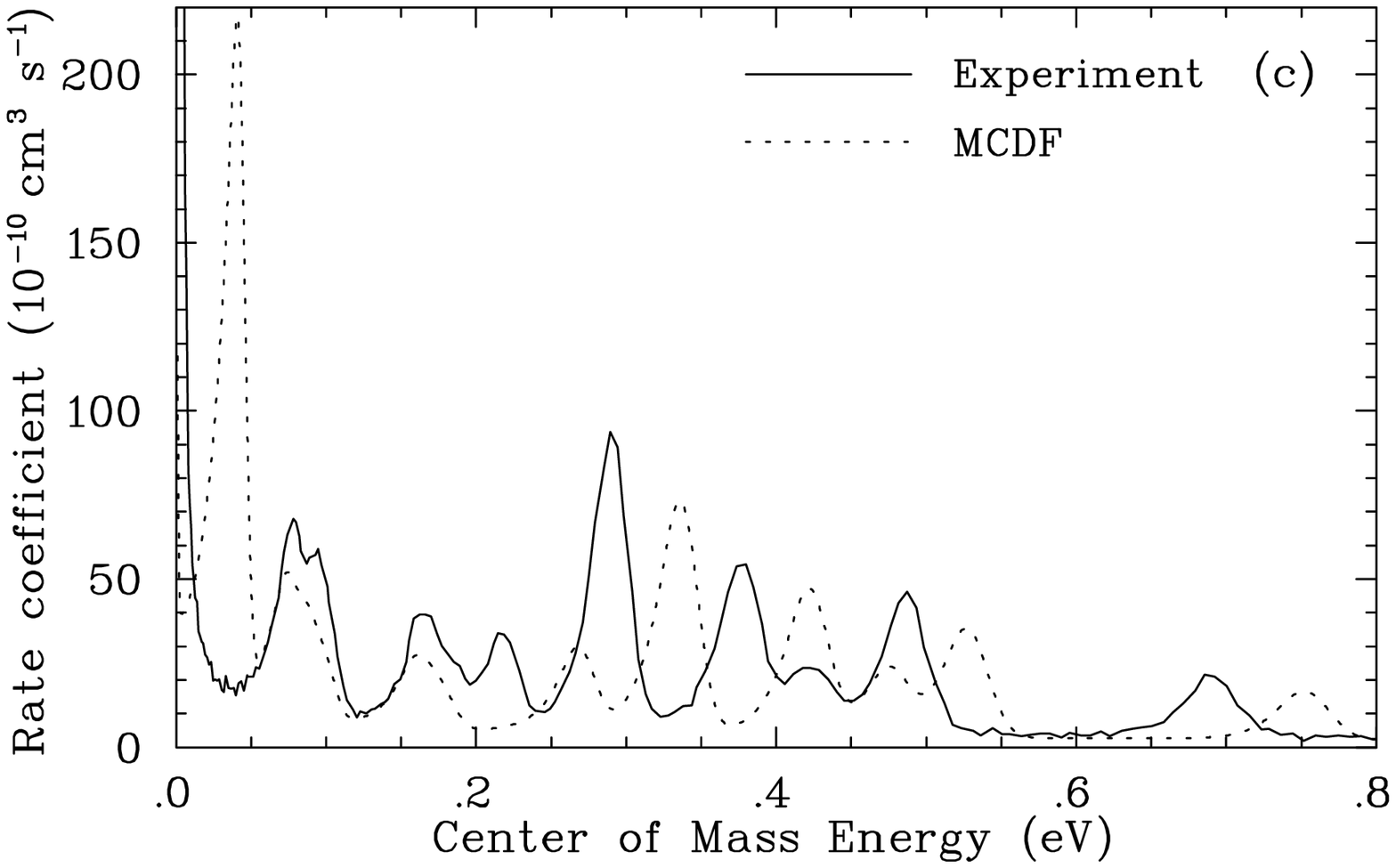}
\caption{\it Continued}
\end{figure}

\setcounter{figure}{4}
\begin{figure}
\plotone{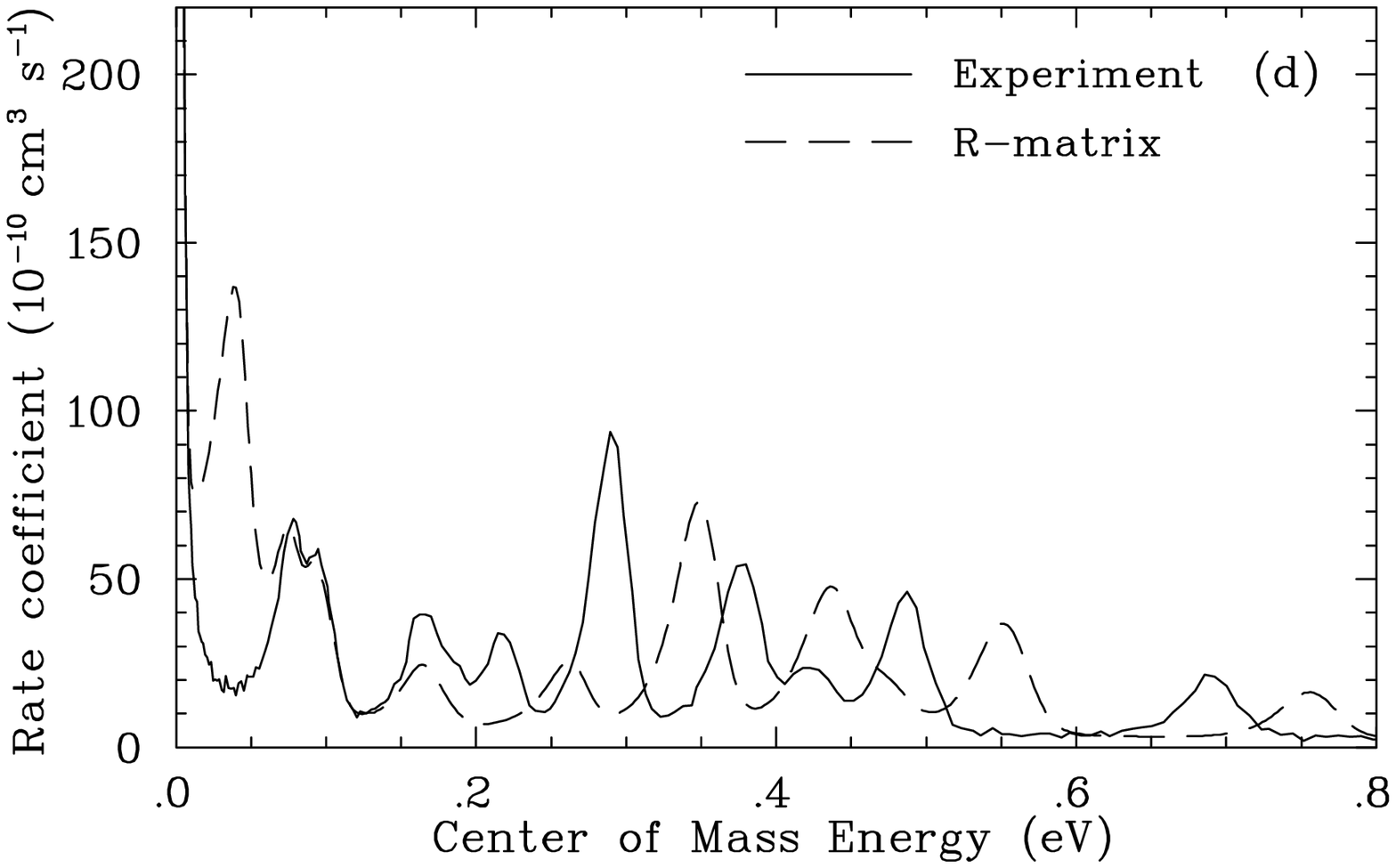}
\caption{\it Continued}
\end{figure}

\begin{figure}
\plotone{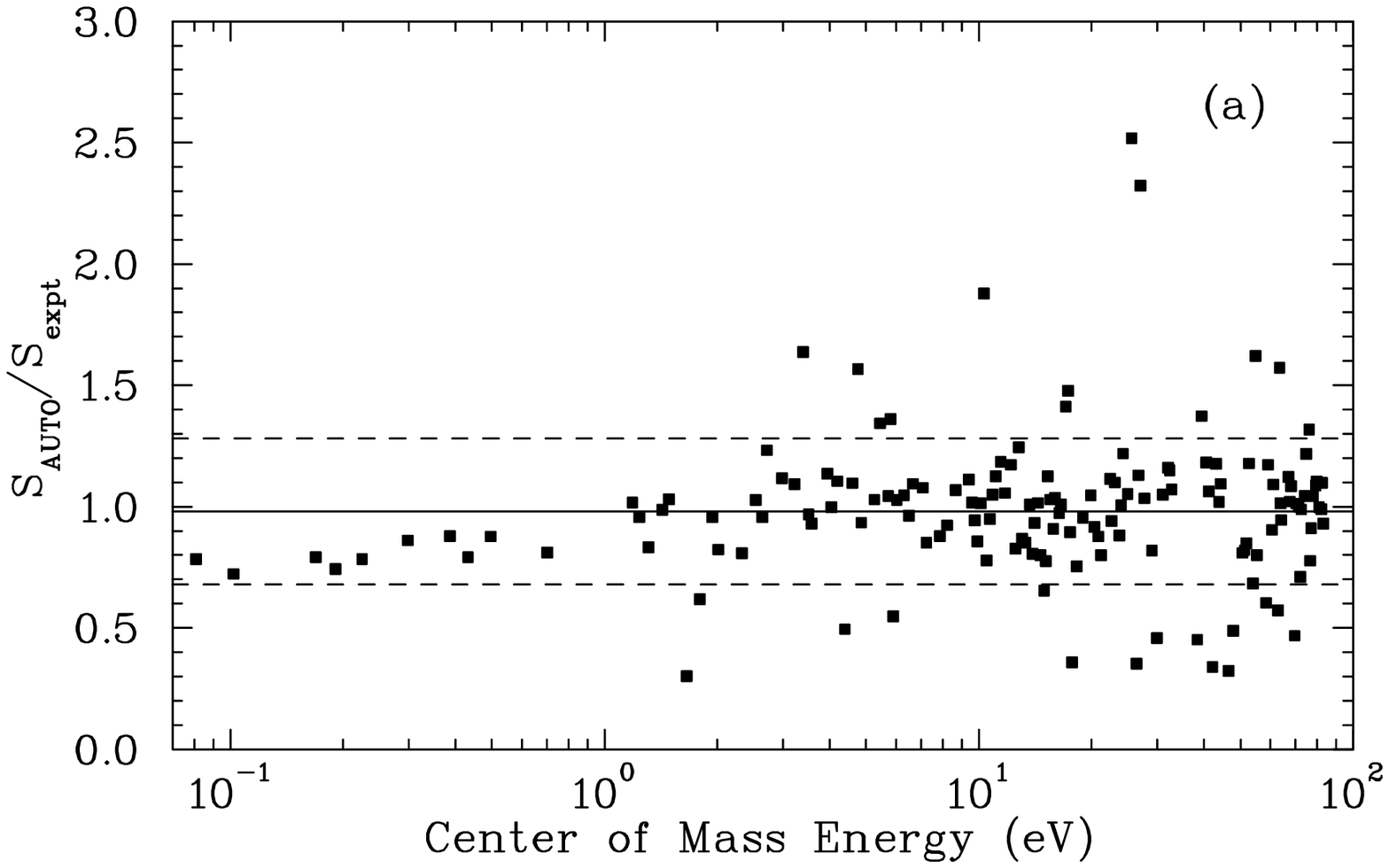}
\caption{The ratio of the resonance strengths given in
Table~\ref{tab:FeXXextracteddata} for our (a) AUTOSTRUCTURE/experiment,
(b) HULLAC/experiment, and (c) MCDF/experiment, results.  
Resonance strength
ratios are shown as a function of center-of-mass collision energy from
0.07 to 100 eV.  The solid lines show the average value for the various
ratios.  The dashed lines show the 1$\sigma$ standard deviation from
these average values.}
\label{fig:strengthratios}
\end{figure}

\setcounter{figure}{5}
\begin{figure}
\plotone{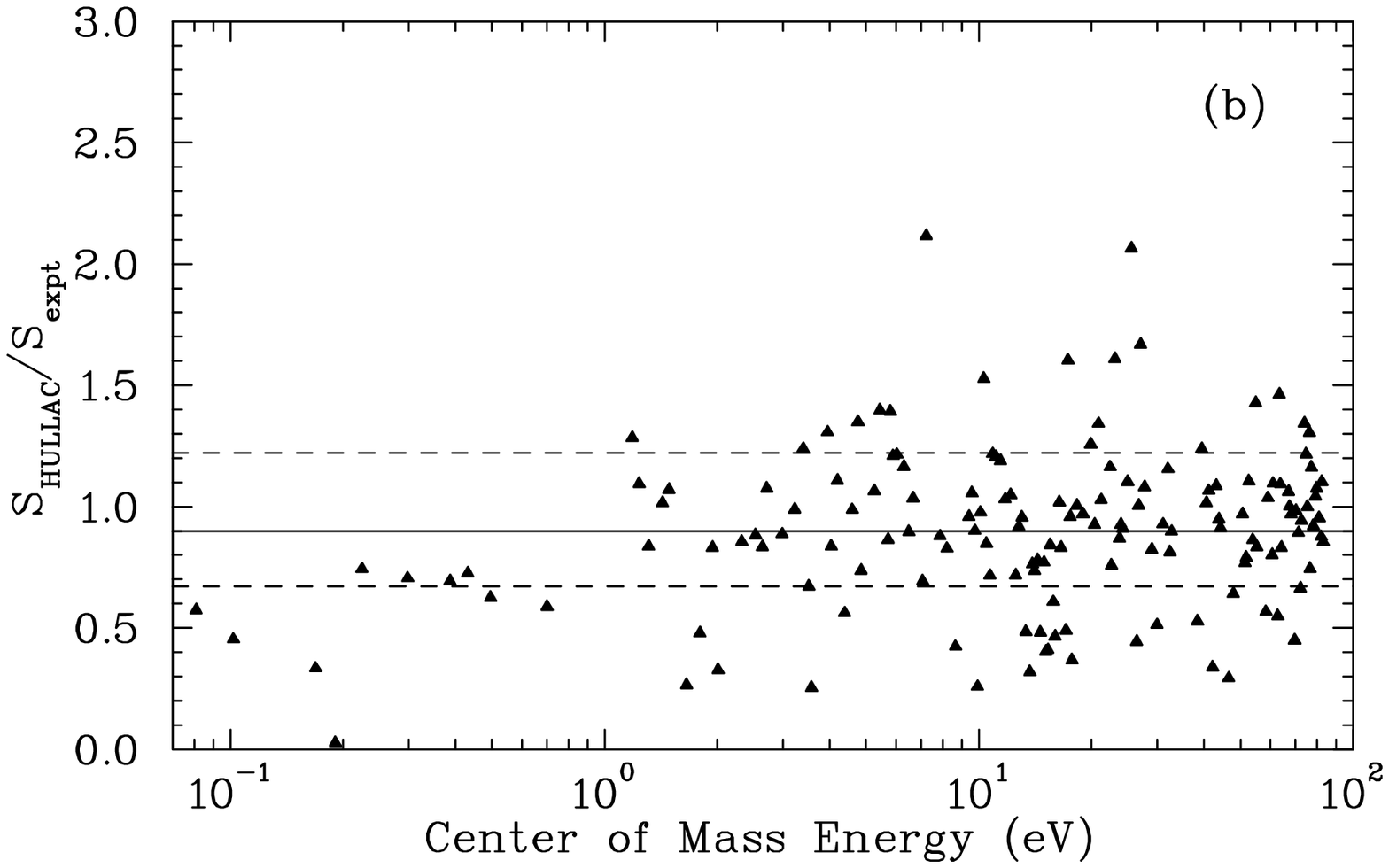}
\plotone{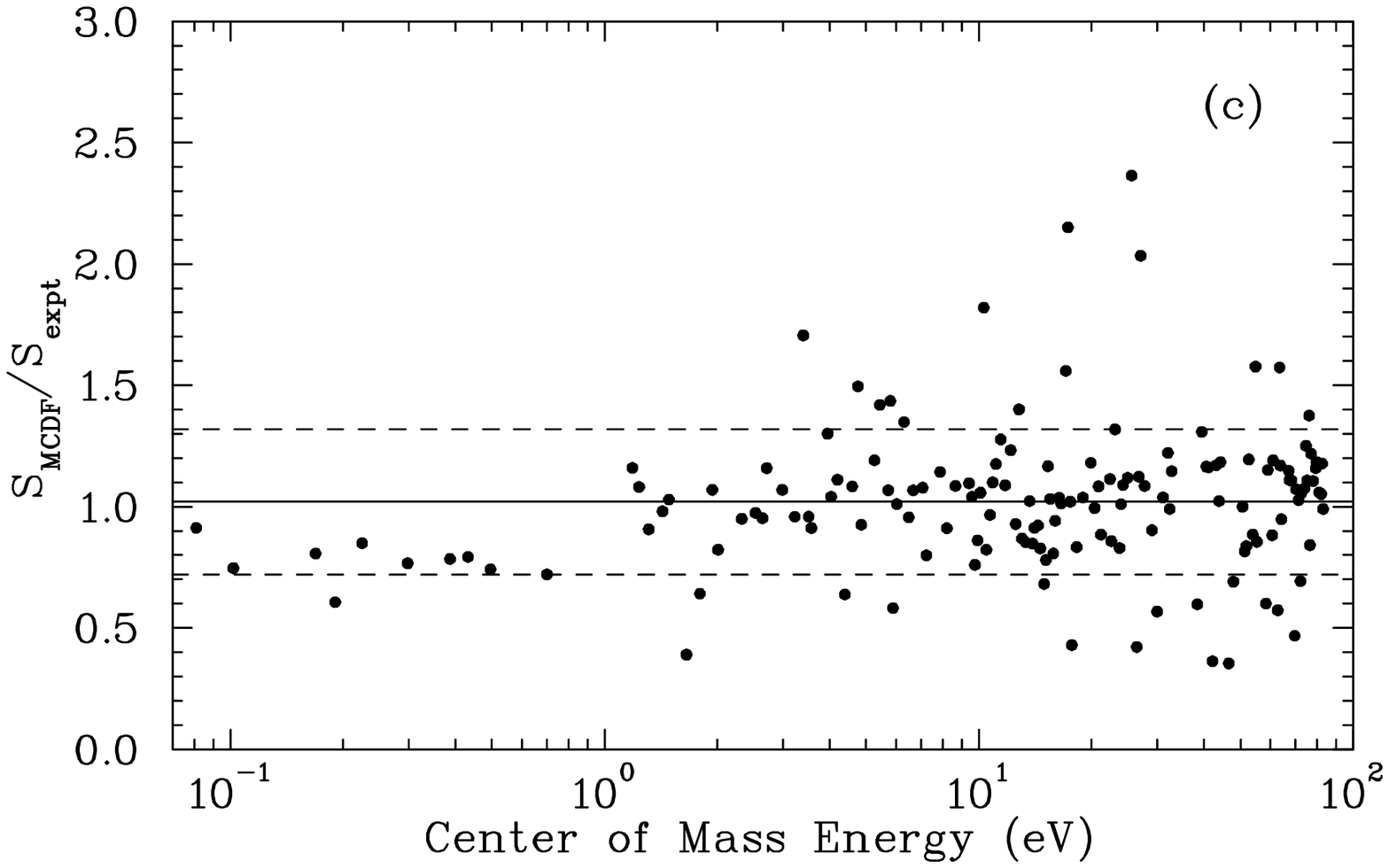}
\caption{\it Continued}
\end{figure}

\begin{figure}
\plotone{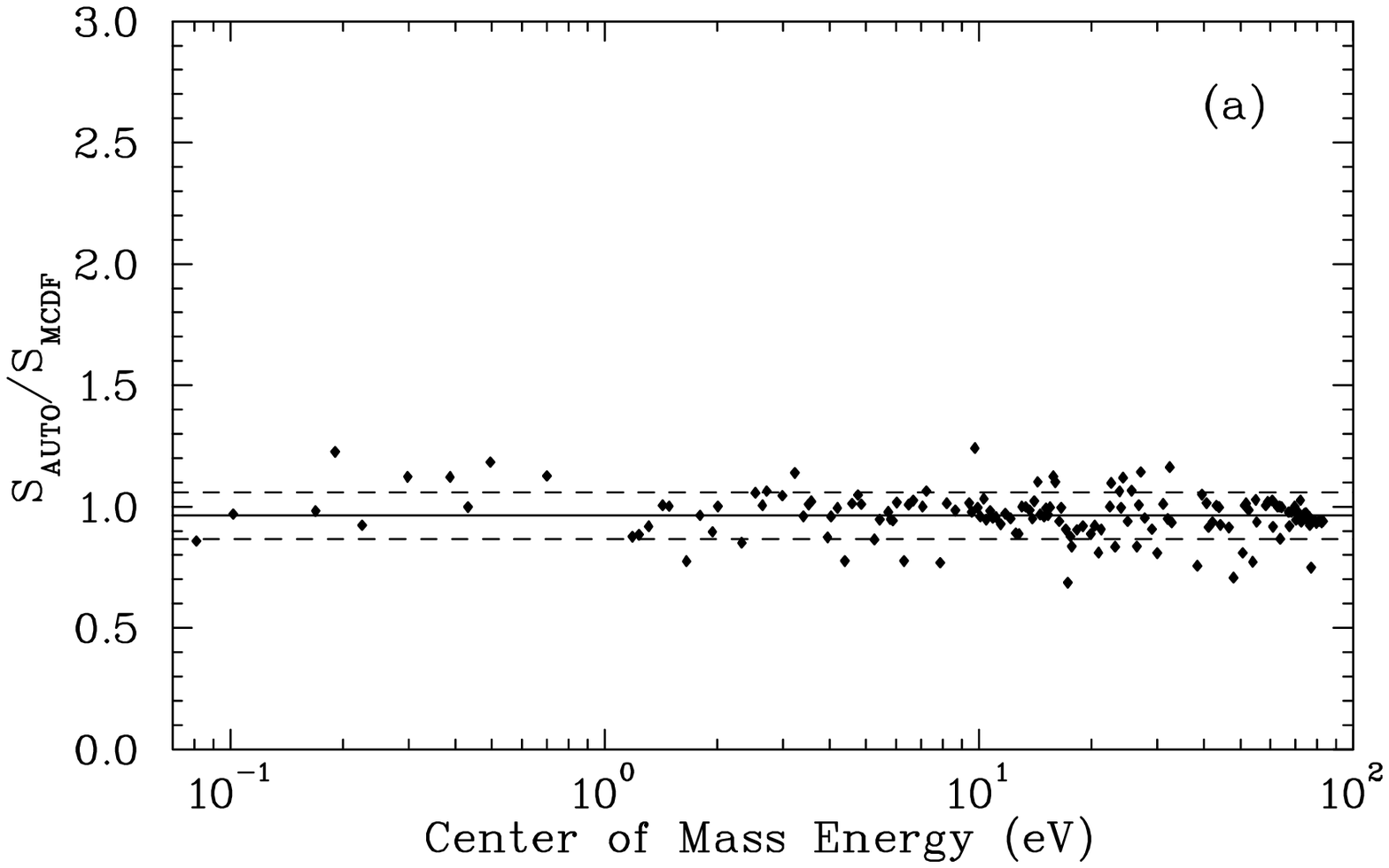}
\caption{The ratio of the resonance strengths given in
Table~\ref{tab:FeXXextracteddata} for our (a) AUTOSTRUCTURE/MCDF, (b)
HULLAC/MCDF, and (c) HULLAC/AUTOSTRUCTURE results.  Resonance strength
ratios are shown as a function of center-of-mass collision energy from
0.07 to 100 eV.  The solid lines show the average value for the various
ratios.  The dashed lines show the 1$\sigma$ standard deviation from
these average values.}
\label{fig:theorystrengthratios}
\end{figure}

\setcounter{figure}{6}
\begin{figure}
\plotone{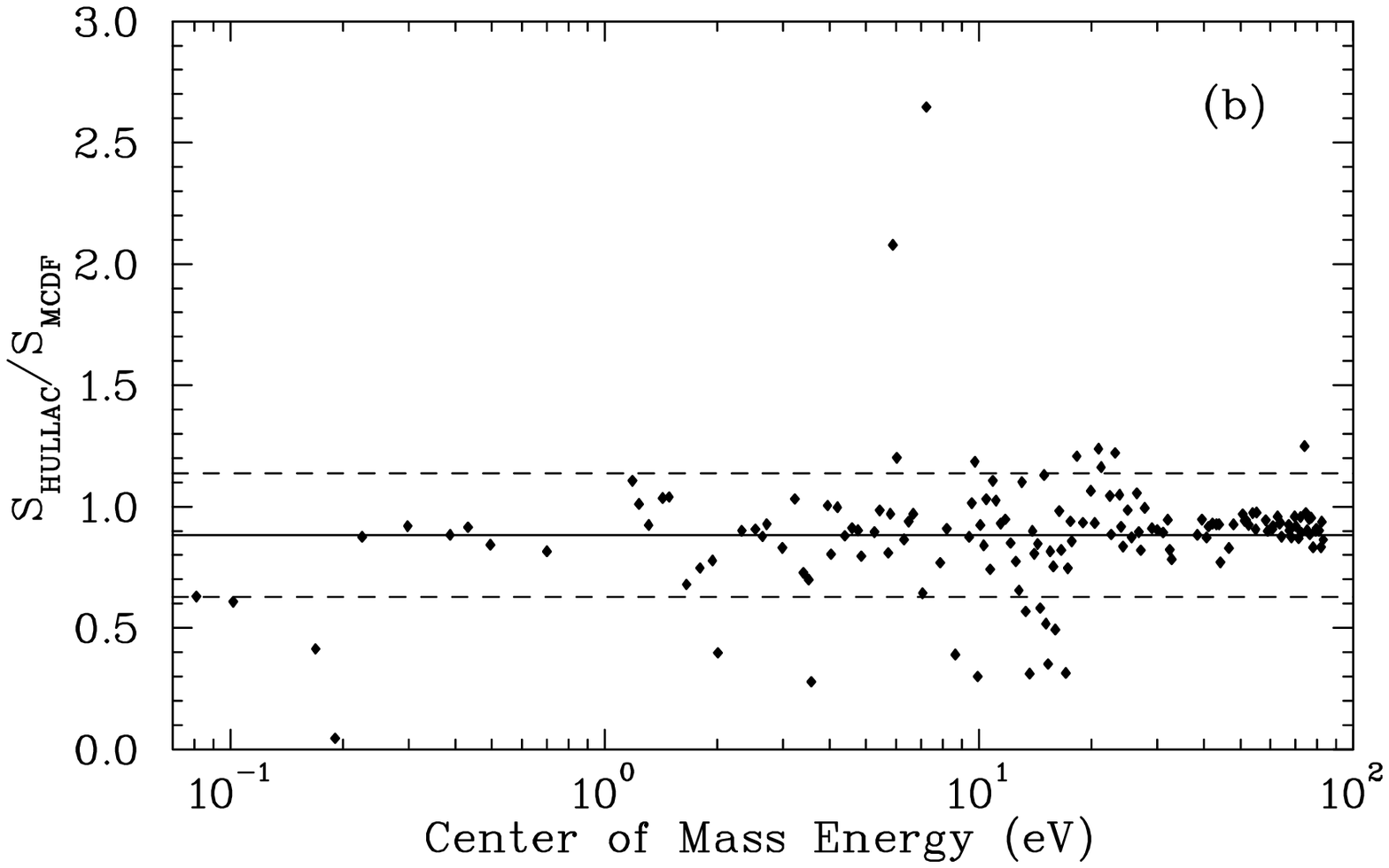}
\plotone{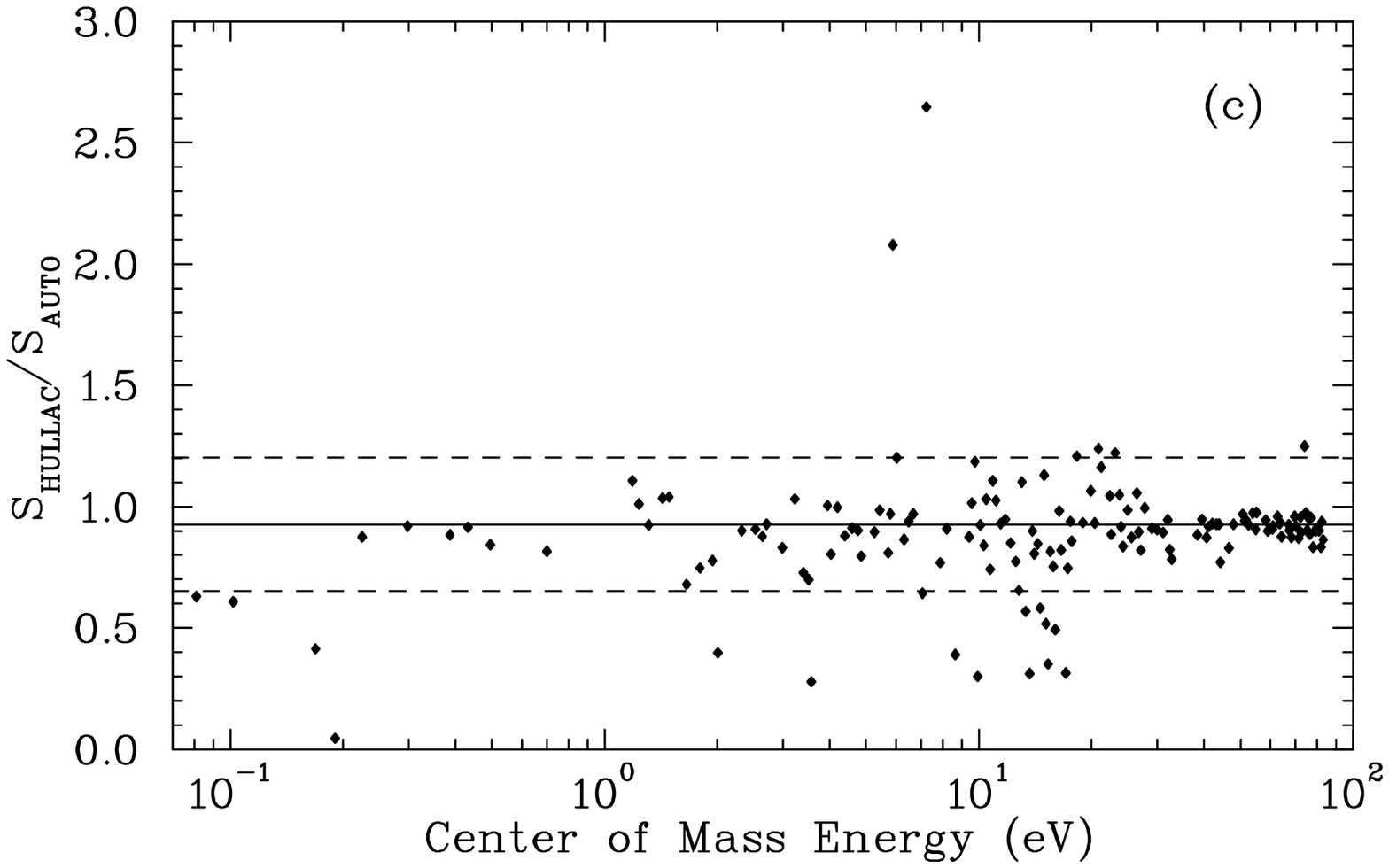}
\caption{\it Continued}
\end{figure}


\begin{thebibliography}{JUNK}

\bibitem [Arnaud \& Raymond(1992)] {Arna92a} Arnaud, M., \& Raymond, J. C.
1992, \apj, 398, 394

\bibitem [Aymar et al.(1996)] {Ayma96a} Aymar, M., Greene, C. H., \&
Luc-Koenig, E. 1996, Rev. Mod. Phys., 68, 1015

\bibitem [Babb et al.(1992)] {Babb92a} Babb, J. F., Habs, D., Spruch,
L., \& Wolf, A. 1992, Z. Phys. D, 23, 197

\bibitem [Badnell(1986)] {bad86} Badnell, N. R. 1986, J. Phys. B, 19, 3827

\bibitem [Bar-Shalom et al.(1988)] {BarS88a} Bar-Shalom, A., Klapisch, M.,
\& Oreg, J. 1988, Phys. Rev. A, 38, 1733

\bibitem [Bar-Shalom(2001)] {BarS01b} Bar-Shalom, A. 2001, private 
communication

\bibitem [Bar-Shalom et al.(2001)] {BarS01a} Bar-Shalom, A., Klapisch,
M., \& Oreg, J. 2001, J. Quant. Spectrosc. Radiat. Transfer, in press 

\bibitem [Behar et al.(1995)] {Beha95a} Behar, E., Mandelbaum, P.,
Schwob, J. L., Bar-Shalom, A., Oreg, J., \& Goldstein, W. H. 1995,
Phys. Rev. A, 52, 3770

\bibitem [Behar et al.(1996)] {Beha96a} Behar, E., Mandelbaum, P.,
Schwob, J. L., Bar-Shalom, A., Oreg, J., \& Goldstein, W. H. 1996,
Phys. Rev. A, 54, 3070

\bibitem [Berrington et al.(1987)] {Berr87a} Berrington, K. A., 
Burke, P. G., Butler, K., Seaton, M. J., Storey, P. J., Taylor,
K. T., and Yan, Y. 1987, J. Phys. B, 20, 6379

\bibitem [Berrington et al.(1995)] {berr95} Berrington, K. A., Eissner, W.
B., \& Norrington, P. H. 1995, Comput. Phys. Commun., 92, 290

\bibitem [Bhatia et al.(1989)] {Bhat89a} Bhatia, A. K., Seely, J. F.,
\& Feldman, U. 1989, At. Data Nucl. Data Tables, 43, 99

\bibitem [Burke \& Berrington(1993)] {burke93} Burke, P. G., \&
Berrington, K. A.\ 1993, Atomic and Molecular
Processes: An R-matrix Approach, (Bristol: IOP Publishing)

\bibitem [Chen(1985)] {Chen85a} Chen, M. H. 1985, \pra, 31, 1449

\bibitem [Cheng, Kim, \& Desclaux(1979)] {Chen79a} Cheng, K. T., Kim,
Y.-K., \& Desclaux, J. P. 1979, At. Data Nucl. Data Tables, 24, 111

\bibitem [Cottam et al.(2001)] {Cott01a} Cottam, J., Kahn, S. M., 
Brinkman, A. C., den Herder, J. W., \& Erd, C. 2001, A\&A, 365, L277

%\bibitem [DeWitt et al.(1995)] {DeWi95a} DeWitt, D. R., Lindroth, E.,
%Schuch, R., Gao, H., Quinteros, T., \& Zong, W. 1995, J. Phys. B, 28, L147

\bibitem [Donnelly et al.(1999)] {Donn99a} Donnelly, D., Bell, K. L., \&
Keenan, F. P. 1999, \mnras, 307, 595

\bibitem[Ferland et al.(1998)]{Ferl98a}Ferland, G. J., Korista, K.
T., Verner, D. A., Ferguson, J. W., Kingdon, J. B., \& Verner, E. M.
1998, PASP, 110, 761

\bibitem [Froese-Fischer(1991)] {ff91} Froese-Fischer, C. 1991,
Comput. Phys. Commun., 64, 369

\bibitem [Gorczyca et al.(1995)] {gorczyca95} Gorczyca, T. W., Robicheaux,
F., Badnell, N. R., \& Pindzola, M. S.\ 1995, Phys. Rev. A, 52, 3852

\bibitem [Gorczyca et al.(1996)] {gorczyca96} Gorczyca, T. W., Robicheaux,
F., Badnell, N. R., \& Pindzola, M. S.\ 1996, Phys. Rev. A, 54, 2107

\bibitem [Gorczyca \& Badnell(1997)] {gorczyca97a} Gorczyca, T. W., \&
Badnell,
 N. R.\ 1997, Phys. Rev. Lett., 79, 2783

\bibitem [Gorczyca et al.(1997)] {gorczyca97b} Gorczyca, T. W., Pindzola,
M. S., Robicheaux, F., \& Badnell,
N. R.,\ 1997, Phys. Rev. A, 56, 4742

\bibitem [Grant et al.(1980)] {Gran80a} Grant, I. P., McKenzie, B. J.,
Norrington, P. H., Mayers, D. F., \& Pyper, N. C. 1980, Comput. Phys.
Commun., 21, 207

\bibitem [Gwinner et al.(2000)] {Gwin00a} Gwinner, G. et al.\ 2000, \prl,
84, 4822

\bibitem [Hess, Kahn, \& Paerels(1997)] {Hess97a} Hess, C. J., Kahn, S. M.,
\& Paerels, F. B. S. 1997, \apj, 478, 94

\bibitem [Hoffknecht et al.(1998)] {Hoff98a} Hoffknecht, A. et al.\ 1998,
J. Phys. B, 31, 2415

\bibitem [Hoffknecht et al.(2001)] {Hoff01a} Hoffknecht, A., Schippers, S.,
M\"uller, A., Gwinner, G., Schwalm, D., \& Wolf. A. 2001,
Physica Scripta, accepted

\bibitem [Jacobs et al.(1977)] {Jaco77a} Jacobs, V. L., Davis, J., Kepple,
P. C., \& Blaha, M. 1977, \apj, 211, 605

\bibitem [Kallman \& Bautista(2001)] {Kall01a} Kallman, T. R. \&
Bautista M. 2001, \apjs, 133, 221

\bibitem[Kaspi et al.(2000)] {Kasp00a} Kaspi, S., Brandt, W. N., Netzer,
H., Sambruna, R., Chartas, G., Garmire, G. P., \& Nousek, J. A. 2000,
\apjl, 535, L17

\bibitem [Kilgus et al.(1992)] {Kilg92a} Kilgus, G., Habs, D., Schwalm,
D., Wolf, A., Badnell, N. R., \& M\"{u}ller, A. 1992, \pra, 46, 5730

\bibitem [Kinkhabwala et al.(2001)] {Kink01a} Kinkhabwala, A. et al.
2001, in preparation

\bibitem [Klapisch(1971)] {Klap71a} Klapisch, M. 1971, Comput. Phys.
Commun., 2, 239

\bibitem [Klapisch et al.(1977)] {Klap77a} Klapisch, M., Schwob, J. L.,
Fraenkel, B., \& Oreg, J. 1977, J. Opt. Soc. Am., 67, 148

\bibitem [Lampert et al.(1996)] {Lamp96a} Lampert, A., Wolf, A., Habs,
D., Kilgus, G., Schwalm, D., Pindzola, M. S., \& Badnell, N. R. 1996,
\pra, 53, 1413

\bibitem [Marxer \& Spruch(1991)] {Marx91a} Marxer, H., \& Spruch, L. 1991,
\pra, 43, 1268

\bibitem [Mitnik et al.(1999)] {Mitn99a} Mitnik, D. M., Pindzola, M. S.,
\& Badnell, N. R. 1999, Phys. Rev. A {\bf 59}, 3592

\bibitem[M\"uller et al.(1987)] {Mull87a} M\"uller, A., et al.\ 1987, \pra,
36, 599

\bibitem[M\"uller \& Wolf(1997)] {Mull97a} M\"uller, A., \& Wolf, A. 1997,
in Accelerator-Based Atomic Physics Techniques and Applications,
ed. S. M. Shafroth \& J. C. Austin, (New York: American Institute of
Physics), 147

\bibitem [Oreg et al.(1991)] {Oreg91a} Oreg, J., Goldstein, W. H.,
Klapisch, M., \& Bar-Shalom, A. 1991, Phys. Rev. A, 44, 1750

\bibitem [Paerels et al.(2000)] {Paer00a} Paerels, F. et al.\ 2000, \apjl,
533, L135

\bibitem [Pindzola et al.(1992)] {Pind92a} Pindzola, M. S., Badnell, N. R.,
\& Griffin, D. C. 1992, \pra, 46, 5725

\bibitem [Price(1997)] {Pric97a} Price, A. D. 1997, Ph.D. Thesis,
University of Strathclyde, UK

\bibitem [Robicheaux et al.(1995)] {robicheaux95} Robicheaux, F., Gorczyca,
T. W., Pindzola, M. S., \& Badnell, N. R. 1995, Phys. Rev. A, 52, 1319

\bibitem [Roszman(1987)] {Rosz87a} Roszman, L. J. 1987, \pra, 35, 3368

\bibitem [Saghiri et al.(1999)] {saghiri1999} Saghiri, A. A.
et al.\ 1999, Phys. Rev. A, {60}, R3350.

\bibitem [Savin(1999)] {Savi99b} Savin, D. W. 1999, \apj, 523, 855

\bibitem [Savin et al.(1997)] {Savi97a} Savin, D. W. et al.\ 1997, \apj, 489,
L115

\bibitem [Savin et al.(1999)] {Savi99a} Savin, D. W., et al.\ 1999, \apjs,
123, 687

\bibitem [Savin et al.(2000)] {Savi00a} Savin, D. W., et al.\ 2000 in
Proceedings of the 12th APS Topical Conference on Atomic Processes in
Plasmas, Reno Nevada, ed. R. C. Mancini (New York: American Institute
of Physics), p.\ 267

\bibitem [Schippers et al.(1998)] {Schi98a} Schippers, S., Bartsch, T.,
Brandau, C., Gwinner, G., Linkemann, J., M\"uller, A., Saghiri, A. A.,
\& Wolf, A. 1998, J. Phys. B, 31, 4873

\bibitem [Schippers et al.(2001)] {Schi01a} Schippers, S., Bartsch, T.,
Brandau, C, M\"uller, A., Gwinner, G., Wissler, G., Beutelspacher, M.,
Grieser, G., \& Wolf, A. 2001, \pra, 62, 022708

\bibitem [Scott \& Taylor(1982)] {scott82} Scott, N. S., \& Taylor, K. T.
1982, Comput. Phys. Commun., 25, 347

\bibitem [Seaton \& Storey(1976)] {Seat76a} Seaton, M. J., \& Storey,
P. J. 1976, in Atomic Processes and Applications, ed. P. G. Burke
\& B. L. Moisewitch (North-Holland, Amsterdam), 133

\bibitem [Seaton(1983)] {seaton83} Seaton, M. J. 1983, Rep. Prog. Phys. 46,
167

\bibitem [Shull \& van Steenberg(1982)] {Shul82a} Shull, J. M., \&
van Steenberg, M. 1982, \apjs, 48, 95; erratum \apjs 49, 351

\bibitem [Sugar \& Corliss(1985)] {Suga85a} Sugar, J., \& Corliss, C.
1985, J. Phys. Chem. Ref. Data, 24, Suppl. 2

\bibitem [Theodosiou et al.(1986)] {Theo86a} Theodosiou, C. E., Inokuti,
M., \& Manson, S. T. 1986, At. Data Nucl. Data Tables, 35, 473

\bibitem [Zhang \& Pradhan(2000)] {Zhan00a} Zhang, H. L., \& Pradhan, A. K.
2000, \mnras, 313, 13

\end{thebibliography}
\end{document}